\documentstyle[epsf]{mn}
\def\mg2{Mg$_2$}
\def\fe{$<$Fe$>$}
\def\etal{{\rm et al.~}}

\begin{document}

\title{Two-Dimensional Line Strength Maps in Three Well-studied Early-Type Galaxies}

\author[Peletier et al.]{Reynier F. Peletier,$^{1,2}$
A. Vazdekis,$^{1,3}$
S. Arribas,$^4$
C. del Burgo,$^4$
B. Garc\'\i a-Lorenzo,$^4$
  \newauthor
C. Guti\'errez,$^4$
E. Mediavilla,$^4$
and F. Prada,$^{5,6}$\\
$^1$ Dept. of Physics, University of Durham, South Road, Durham DH1 3LE,
UK \\
$^2$ School of Physics and Astronomy, University of Nottingham, 
Nottingham NG7 2RD, UK \\
$^3$ Institute of Astronomy, University of Tokyo, Japan \\
$^{4}$Instituto de Astrof\'\i sica de Canarias,E-38200 La Laguna,
Tenerife, Spain \\
$^5$UNAM, Ensenada, Mexico \\
$^6$ Calar Alto Observatory, Almeria, Spain}

\maketitle

\markboth{Peletier et al.}{TWO-DIMENSIONAL LINE STRENGTHS}


\begin{abstract}

Integral field spectroscopy has been obtained for the nuclear regions of
3 large, well-studied, early-type galaxies. From these spectra 
we have obtained line strength maps for about 20 absorption lines, mostly
belonging to the Lick system. An extensive comparison with multi-lenslet
spectroscopy shows that accurate kinematic maps can be obtained, and also
reproducible line strength maps. Comparison with long-slit spectroscopy 
also produces good agreement.

We show that Mg is enhanced with respect to Fe in the inner disk of one
of the three galaxies studied, the Sombrero. [Mg/Fe] there is
larger than in the rest of the bulge. The large values of Mg/Fe in
the central disk are consistent with the centres of other early-type
galaxies, and not with large disks, like the disk of our Galaxy, where
[Mg/Fe] $\sim$ 0.  We confirm with this observation  
a recent result of Worthey (1998) that Mg/Fe is determined by  the central
kinetic energy, or escape velocity, of the stars, only, and not by the
formation time scale of the stars.

A stellar population analysis using the models of Vazdekis et al. (1996) 
shows that our observed H$\gamma$ agrees well with what is predicted based 
on the other lines. Given the fact that H$\beta$ is often contaminated 
by emission lines we confirm the statement of Worthey \& Ottaviani (1997),
Kuntschner \& Davies (1998) and others that if one tries to measure ages 
of galaxies H$\gamma$ is a much better index to use than H$\beta$. 
Using the line strength of the Ca II IR triplet as an indicator of the 
abundance of Ca, we find that Ca follows Fe, and not Mg, in these galaxies.
This is peculiar, given the fact that Ca is an $\alpha$-element.
Finally, by combining the results of
this paper with those of Vazdekis et al. (1997) we find that the 
line strength gradients in the three galaxies are primarily caused by variations
in metallicity.
\end{abstract}


\section{Introduction}

The study of the stellar populations of local galaxies plays a major role in
understanding galaxy formation and evolution, not in the least because we
can only understand high-redshift
galaxies by comparing them to the local universe. 
In recent years significant progress has been made in understanding
the stellar populations of nearby galaxies, for ellipticals as well as spirals.
Elliptical galaxies are found to fit well single-age, single-metallicity (SSP)
models (e.g. Bruzual \& Charlot 1993, Worthey 1994, Vazdekis \etal 1996), 
suggesting that most stars that we currently see were formed on a small time
scale. Large ellipticals generally have non-solar Mg/Fe ratios 
(Peletier 1989, Worthey \etal 1992), while it appears that there is a large
spread in age among them (Gonz\'alez 1993), although this is not generally
accepted (see Kuntschner \& Davies 1998). Spiral galaxies have been less
well studied, because of the presence of extinction and star formation makes
stellar population studies much more difficult. Recent reviews about
the stellar populations in bulges, and their differences with 
ellipticals, are, e.g., Worthey (1998) and Peletier (1999).

This paper is part of a series of papers, with the aim of studying galactic 
bulges in the same detail as ellipticals. To do this, we have first developed 
a new spectrophotometric stellar population model, that can be used to interpret
observed colours and absorption lines of galaxies (Vazdekis \etal 1996, 
hereinafter Paper I). To test this model, we have presented in a subsequent 
paper high-quality long-slit observations of three representative early-type
galaxies, in many lines and colours, and applied the model to it. This allowed
us to obtain a detailed understanding of ages and abundance distributions 
of these well-studied, 'standard', galaxies (Vazdekis \etal 1997, hereinafter Paper II).
In the current paper, we again present the same three galaxies (NGC 3379,
NGC 4472 and NGC 4594) but now using two-dimensional data, derived using 
Integral Field Spectroscopy (IFS). This is the first time a detailed study
has been performed of galaxy absorption line strengths using a multi-fiber
instrument. A similar, but much more limited 
study has been performed by Emsellem \etal
(1996), who used the multi-lenses IFU system TIGER on the CFHT to obtain
2-dimensional kinematic and line strength maps of the Sombrero galaxy. Although
the number of absorption lines studied by them is much smaller, 
their paper offers 
us an excellent opportunity to carefully check our IFU results. Two-dimensional
spectroscopy of early-type galaxies has also been obtained by
Sil'chenko and collaborators (e.g. Sil'chenko et al. 1997, 1999), but
in their papers only two-dimensional kinematics has been presented.

There are several advantages in using IFS as opposed to multiple long slit
spectra. Once the problems of data reduction have been overcome, the larger
spatial coverage will be a great advantage in studying is much the galaxy
kinematics. For example, one can study the triaxiality of a system  much better
than using long slits (see e.g. Statler 1991). Studying e.g. whether
photometric and kinematic centres coincide is now much easier. With regard to
stellar populations: one can find two-dimensional stellar population 
structures, like young disks and remnants of mergers, purely by analyzing line
strength maps. One can detect small amounts of star formation everywhere in the
area covered. And, compared to long-slit spectroscopy, the sensitivity at a 
certain radius is much larger, since more area is covered. We will discuss in
this paper results obtained using a fiber bundle of 125 fibers. In the  near
future however, thousands of fibers will be used, which undoubtedly will  make
IFS much more popular and important than it is at present.

In this paper, several tests will be presented  to establish our IFS reduction
techniques. But also, we will present some entirely new data, for the H$\gamma$
line, and for the Ca II IR triplet. H$\gamma$ is a good age-indicator, when
used together with e.g. Mg$_2$ (Worthey \& Ottaviani 1997), and can be used 
even when H$\beta$ is filled in by emission. The Ca II IR triplet in the 
near-infrared is one of the few lines depending strongly 
on stellar surface gravity
(see e.g. Carter \etal 1987), allowing us to possibly measure or constrain
the IMF slope. In this paper we present the Ca triplet (which we will address
as Ca T in the rest of this paper) calibrated using stars
of D\'\i az \etal (1989).

We will present very little kinematics here, only a comparison with 
Emsellem \etal (1996), to show the reliability of our velocity dispersion
corrections. Kinematics, including line profile analysis, will be presented 
in a future paper (Prada et al. 1999, in preparation). 
Here we present stellar population 
analysis of the 3 well-studied galaxies. At the same time, we also observed
2 other galaxies: the dwarf elliptical M~32 and the Sb NGC~2841. The 
analysis of these two galaxies will also be presented in future papers.
The organization of our paper is as follows. In Section 2 the observations
are described. In Section 3 we described the data reduction. In Section 
4 a comparison is made with Paper II and Emsellem \etal (1996). In
Section 5 we present our results, amongst other the population synthesis.
In the last Section our conclusions are given. 

\section{Observations}

\subsection{The Instrument}

The data analyzed in this paper were obtained on Feb. 15-17, 1997, at the
Observatorio del Roque de los Muchachos, on the
island of La Palma. We used the 2D-FIS (Two-dimensional Fiber ISIS System),
which is placed between the f/11 Cassegrain focus of
the 4.2 m William Herschel Telescope (WHT) and the ISIS double spectrograph. A
detailed technical description is
provided by Garc\'\i a \etal (1994); here we will only recall its main
characteristics. The core of the system consists of a 2.5 m
long bundle formed by 125 optical fibers, each 200 $\mu$m (0.9'' on the sky) in
diameter, arranged in two groups in the focal
plane. One has 95 fibers, forming an array 9.4'' $\times$ 12.2'' on the sky, 
while the other, a ring of 38'' in radius with 30 fibers, is 
intended to collect the background light.  
The relative positions of the fibers at the
telescope's focal plane are known very accurately, the distance between two
adjacent fibers being about 1.1''. Figure 1
shows the actual distribution of the fibers in the central rectangle. At the
slit the fibers have been arranged linearly (with a
distance of 425 $\mu$m between adjacent fibers). 
With this type of arrangement, a set
of spectra - corresponding to 125 zones
in the circumnuclear region - is recorded in each exposure. 
The advantage of the system as opposed to, say, a Fabry-Perot
is that all the spectral and spatial information is recorded simultaneously and
that the spectral coverage is determined 
by the spectrograph alone. Details on the technique itself may be found in
Arribas, Mediavilla, \& Rasilla (1991), and
references therein.

\begin{figure}
\mbox{\epsfxsize=8cm  \epsfbox{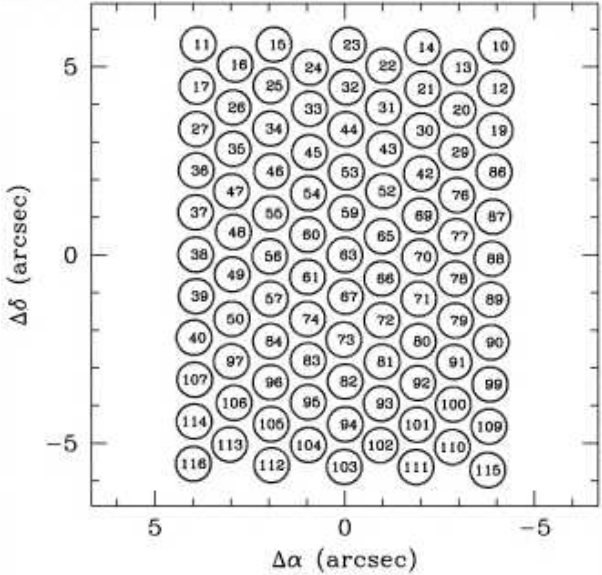}}
\caption{Spatial distribution of the fiber faces of the central rectangle of the
optical fiber bundle, at the 4.2 m WHT focal
plane. The central fiber (labeled 63) is the
origin of the coordinate axes. North is at the top and east is to the left.
The numbers reflect the order of the fibers on the entrance slit.}
\end{figure}

\subsection{Our Observations}

The observations were made using simultaneously the blue and red arms of the
ISIS double spectrograph. In the blue arm, 
the R300 grating was used, giving us a
pixel scale of 96 km/s (1.55 ${\rm \AA}$), a range between 4100 and 5650 
${\rm \AA}$, covering
most of the lines of the Lick system, including H$\gamma$  (Worthey 
\& Ottaviani 1997), and a resolution of 2 pixels. In the red, we centered 
our spectrum on the Ca T. Here we had a pixel size of 28 km/s
(resolution 2 pixels) and a spectral range from 8200 to 9000 ${\rm \AA}$.
On both arms 1024 $\times$ 1024 thinned TEK CCDs were used. A
dicroic, centered at 6100 ${\rm \AA}$, was used, and in the red arm a 
GG495 order sorting filter was placed as well. Weather conditions during
the first two nights, during which the three galaxies were observed, were
photometric, with a seeing of $\approx$ 1''.
A log of the observations is presented in Table 1.

\begin{table}
\begin{center}
\caption{Observation Log}
\begin{tabular}{lcc}
\hline
\hline
Galaxy & Date & total t$_{\rm int}$   \\
\hline
NGC 4594 (Sombrero) & 15 Feb & 9000s \\
NGC 3379 & 16 Feb & 10800s \\
NGC 4472 & 16 Feb & 7200s \\
\hline 
\end{tabular}
\end{center}
\end{table}

Each night, we also observed some out-of-focus standard stars, using the 
same setup. In total 20 stars were observed.

\section{Data Reduction}

Since observations of this type are not very common, and some steps
of the data reduction process are significantly different from reducing 
e.g. long-slit data, we will cover the reduction process in quite some
detail, and describe a number of tests showing that our results are
reliable. To summarize, we first extract the light in the individual 
galaxy fibers after removing the stray light between the fibers. We
then correct each fiber for its relative transmission using a twilight
sky image, and correct for pixel to pixel sensitivity variations using
tungsten lamp flatfields. After this we calibrate the spectra in wavelength
using CuAr arc lamp exposures, remove cosmic rays, and subtract the 
sky background. We then extract for each fiber absorption line indices
as well as kinematic information, and make maps of these. All of this 
work was done in IRAF, with some programs written by ourselves. Here we 
describe these steps in more detail.

\subsection{Stray Light}

The first reduction step was to remove the stray light. This light has a diluting
effect on the spectra, making it more difficult to measure line indices of the
object. For 2D-FIS  the individual fibers are well separated along the spatial
direction of the CCD detector: the distance between adjacent fibers is 6 pixels,
and the aperture width is 4 pixels. In Fig.~2 a cut along the spatial direction
shows the relative separations of the fibers. For this set up, the optical
cross-talk (contamination on a fiber due to the wings of the nearer fibers) is
negligible if the difference in intensity between adjacent fibers 
is not too large. This is the case for most objects, except in focus standard
stars. The object images are not very much affected by cross-talk 
because the fibers are conveniently ordered along the spectrograph slit. 
We determined the stray light to be subtracted by fitting legendre polynomia 
through the spatial regions between the fibers, and subtracted it.
Subtracting stray light however barely affects the final line indices. 
This is mostly due to the
fact that the galaxy light across the frame doesn't vary very much, so that an
addition of a few percent of light from elsewhere in the galaxy will not affect
the integrated line index. Experimentally we found that for the Sombrero in the 
blue the average values of the line indices did not change by more  
than 0.2 ${\rm
\AA}$. The RMS fiber difference between correcting and not correcting for stray
light was about 0.3 ${\rm \AA}$ per index (0.005 for Mg$_1$ and Mg$_2$), of the
same order as the numbers in Table 3, indicating that by correcting for stray 
light we might improve the data. For that reason the correction has been 
applied, although it remains uncertain whether it was really necessary.

\begin{figure}
\mbox{\epsfxsize=9cm  \epsfbox{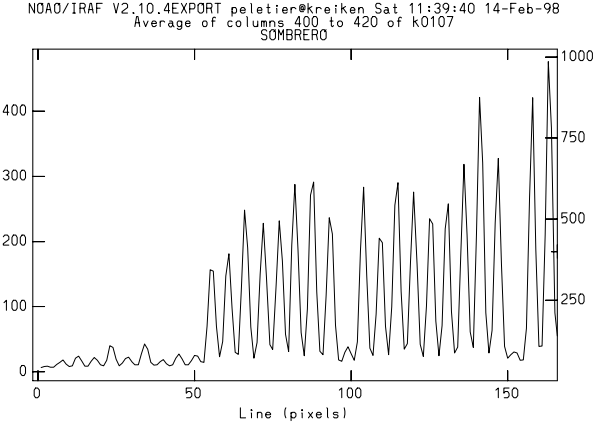}}
\caption{First $\sim$ 30 fibers along the slit in the blue of the Sombrero
galaxy, 
showing the positions of the individual fibers. The first 9 fibers belong
to the outer ring to measure the sky background.}
\end{figure}

\subsection{Flatfielding and Extracting the Fibers}

Flatfielding was done to remove sensitivity variations from pixel to pixel of the
system, and fiber to fiber transmission variations. The TEK CCDs that we used
suffer considerably from fringing in the red (Peletier 1996), but also in the
blue, all the way up  to 5000 ${\rm \AA}$. We measure fringing amplitudes of about
10\%  in our  whole red region and between 4200 and 4400 ${\rm \AA}$, and about
5\% between 4400 and 5000 ${\rm \AA}$. As explained in the technical note, this
fringing depends only on the position of the telescope, and can be corrected for
very well. For measuring maps of very tiny absorption lines however, our final
accuracy might be determined by how well we can correct for this fringing. At the
position of each object, star and galaxy, we also measured  a Tungsten flatfield,
and used this to correct for the fringing. The flatfield image was corrected for
stray light, after which the  fibers spectra were extracted. To the extracted
spectra polynomials (10th order in the blue and 3rd order in the red) were fitted,
in order to correct for the spectral response of the Tungsten  lamp, but not for
the fringes. Following this the object fibers (galaxy or star) were extracted
using the flatfield image as a tracer.  This is justified because flat and object
(and arc) images are observed at the same position of telescope and then the
shifts of each fiber are negligible. They were then divided by the normalized
flatfield, and following that each fiber was divided by its relative transmission,
determined from twilight sky frames taken at the beginning and end of each night. 

\subsection{Wavelength Calibration, Cosmic Ray and Background Removal}

Wavelength calibration was done using CuAr arc lamp exposures taken 
before or after each object frame. For each fiber separately the 
corresponding arc spectrum was extracted, and low order polynomials
were fitted to pixel positions of the arc lines as a function of 
wavelength. On the average the accuracy of the fit for an arc line
was 0.08 ${\rm \AA}$ (4.8 km/s) in the blue and 0.05 ${\rm \AA}$ (1.7 km/s) 
in the red. Using these polynomials the spectra were rebinned in
log(lambda). Since for all of our galaxy we had observed at least
4 data frames of 1800s each, we could combine them and remove the 
cosmic rays with a simple median filtering technique. For the stars,
for which only one exposure was taken, very few cosmic rays were 
present because of their short exposure times, and these were removed
using a sigma-clipping algorithm. We then proceeded to subtract the 
sky background. Two methods were tested. We took separate sky exposures
before and after the galaxy exposures. These were scaled to the galaxy
integration time and subtracted. The second method was to take all the
fibers in the outer ring together to determine an average sky spectrum,
and to subtract that from the galaxy fiber spectra. The first method
suffers from the fact that the sky varies with time, so that the sky
spectra have to be scaled in some way. Using the second method one 
has the problem that the outer ring fibers still have some galaxy contribution
in them. In our worst case, in NGC 4472, at 38'' the galaxy is about
15 $\times$ fainter than at 10'' (Peletier \etal 1990), so some 
contamination of the spectra in the outer regions can be expected. However
in these three objects colour and line strength gradients are so 
small (Peletier \etal 1990, Paper II, Gonz\'alez 1993) 
that such a small contamination of max. 6\% will not be noticed. In 
the blue the sky background in the continuum 
is so low compared to the galaxies that 
no trace of it was left after subtracting. In the red the sky subtraction
also was very precise, because the continuum flux of the outer parts of
the galaxies was always at least 8 times larger than the background. In
the region of the Ca II IR triplet there are no sharp emission features 
in the sky background, only a broad emission line from O2 (Nelson 
\& Whittle 1995).

\subsection{Making Maps and Datacubes}

On the now reduced spectra we continued to measure absorption line
strengths. This was done by first determining the stellar radial
velocity and velocity dispersion of each fiber, deredshifting it,
measuring equivalent widths as explained e.g. in Paper II 
and Worthey (1994), calibrating these equivalent widths 
to put them onto the Lick system in the blue (Faber \etal 1985,
Gorgas \etal 1993,
Worthey \etal 1994) and the system of D\'\i az \etal (1989) in the red,
and making two-dimensional maps of each of them.

Velocities and velocity dispersions were determined using
Bender's (1990) Fourier Quotient method. A comparison with 
the literature for the Sombrero galaxy is given in the following
section. After deredshifting the spectra using these velocities, 
indices were determined,
which then had to be corrected for velocity broadening. To calibrate 
this velocity dispersion correction we used our 20 stars (for each star
the average of all fibers was taken), and determined for each index
a velocity dispersion correction curve (as described in 
Paper II). The same was done in the blue and in the red.
In Table 2 our index definitions are given.
We found that the Ca T indices
were affected very much by the fact that the two continuum bands 
were very far from the feature, so that the line strengths of the features
are severely affected by small variations in the continuum. For that
reason we also tried a few different definitions for the 3 lines
of the triplet, from studies of galactic globular clusters. Three
other definitions were adopted, from Armandroff \& Zinn (1988),
Armandroff \& da Costa (1991) and Rutledge et al. (1997) (see Table 2). 

\onecolumn
\begin{table}
\begin{center}
\caption{Index definitions}
\begin{tabular}{lccccccl}
\hline
\hline
Name & \multicolumn{2}{c}{Left Cont.} & \multicolumn{2}{c}{Band} 
& \multicolumn{2}{c}{Right Cont.} & Ref. \\
\hline
\multicolumn{8}{c}{Blue} \\
\hline
Ca 4227 & 4211.000 &4219.750 &4222.250 &4234.750 &4241.000 &4251.000  &Lick \\
G 4300  & 4266.375 &4282.625 &4281.375 &4316.375 &4318.875 &4335.125  &Lick \\
Fe 4383 & 4359.125 &4370.375 &4369.125 &4420.375 &4442.875 &4455.375  &Lick \\
Ca 4455 & 4445.875 &4454.625 &4452.125 &4474.625 &4477.125 &4492.125  &Lick \\
Fe 4531 & 4504.250 &4514.250 &4514.25  &4559.25  &4560.50  &4579.25   &Lick \\
C2 4668 & 4611.500 &4630.250 &4634.00  &4720.25  &4742.75  &4756.50   &Lick \\
H$\beta$& 4827.875 &4847.875 &4847.875 &4876.625 &4876.625 &4891.625  &Lick \\
Fe 5015 & 4946.500 &4977.750 &4977.75  &5054.00  &5054.00  &5065.25   &Lick \\
Mg b    & 5142.625 &5161.375 &5160.125 &5192.625 &5191.375 &5206.375  &Lick \\
Fe 5270 & 5233.150 &5248.150 &5245.65  &5285.65  &5285.65  &5318.15   &Lick \\
Fe 5335 & 5304.625 &5315.875 &5312.125 &5352.125 &5353.375 &5363.375  &Lick \\
Fe 5406 & 5376.250 &5387.500 &5387.50  &5415.00  &5415.00  &5425.00   &Lick \\
Mg1     & 4895.125 &4957.625 &5069.125 &5134.125 &5301.125 &5366.125  &Lick \\
Mg2     & 4895.125 &4957.625 &5154.125 &5196.625 &5301.125 &5366.125  &Lick \\
H$\gamma$a& 4283.50  &4319.75  &4319.75  &4363.50  &4367.25  &4419.75   &WO97 \\
H$\gamma$f& 4283.50  &4319.75  &4331.25  &4352.25  &4354.75  &4384.75   &WO97 \\
\hline
\multicolumn{8}{c}{Red} \\
  Ca 1      &	8447.5 & 8462.5 & 8483 & 8513 & 8842.5 & 8857.5 & Diaz et al. 1989\\
  Ca 2      &	8447.5 & 8462.5 & 8527 & 8557 & 8842.5 & 8857.5 &Diaz et al. 1989\\
  Ca 3      &	8447.5 & 8462.5 & 8647 & 8677 & 8842.5 & 8857.5 &Diaz et al. 1989\\
  CaAZ1     &	8474.0    & 8489.0    & 8490.0    & 8506.0    & 8521.0  & 8531.0    &Armandroff \& Zinn 1988\\
  CaAZ2     &	8521.0    & 8531.0    & 8532.0    & 8552.0    & 8555.0  & 8595.0    &Armandroff \& Zinn 1988\\
  CaAZ3     &	8626.0    & 8650.0    & 8653.0    & 8671.0    & 8695.0  & 8725.0    &Armandroff \& Zinn 1988\\
  CaAD1     &	8474.0    & 8489.0    & 8532.0    & 8552.0    & 8559.0  & 8595.0    &Armandroff \& Da Costa 1991\\
  CaAD2     &	8626.0    & 8647.0    & 8653.0    & 8671.0    & 8695.0  & 8754.0    &Armandroff \& Da Costa 1991\\
  CaTP1     &	8346.0    & 8489.0    & 8490.0    & 8506.0    & 8563.0  & 8642.0    &Rutledge et al. 1997\\
  CaTP2     &	8346.0    & 8489.0    & 8532.0    & 8552.0    & 8563.0  & 8642.0    &Rutledge et al. 1997\\
  CaTP3     &	8563.0    & 8642.0    & 8653.0    & 8671.0    & 8697.0  & 8754.0    &Rutledge et al. 1997\\
\hline 
\end{tabular}
\end{center}
\end{table}
\twocolumn

Due to the fact that neither the stars on the Lick and D\'\i az systems
are flux calibrated, nor our own spectra, we had to put our observations onto
these systems using standard stars. To do this, we took maps of out-of-focus
stars,measured their average line indices, and determined for each index
a linear relation between the literature index and our measurements. In the 
red only a single offset was sufficient. 
In the blue our sensitivity curve is considerably different from that
of the Lick IDS, used to define the Lick system, and therefore the 
conversions are somewhat more complicated. The calibration relations are 
given in Table~3. The RMS scatter
now indicates how accurate our indices are in absolute terms. As one 
can see from Table 3, the RMS scatter is similar to the scatter for 
the Lick system itself (see Worthey \etal 1994, Table 1), which implies that
the absolute values of our line indices are probably better than those
of the Lick system. 

\begin{table}
{\small
\begin{center}
\caption{Conversion to the Lick/Diaz system. Coefficients are given 
of the equation O = a + b C, with O the observed index and C the 
calibrated index. There errors in a and b are derived from the dispersion 
in the stars. $\sigma_{RMS}$ is the RMS dispersion in this relation for
a line strength measurement of an individual star.
In the last column the RMS dispersion from fiber to fiber in each line
index, from observations of defocussed stars, is given.}
\begin{tabular}{lcccccc}
\hline
\hline
Band & a & $\pm$ & b & $\pm$ & $\sigma_{RMS}$ & $\sigma_{FF}$ \\
\hline
Ca 4227   &  0.19   &  0.08      & 1.40 & 0.04  & 0.17  &  0.260  \\
G 4300    &  0.55    & 0.16	 & 1.05 & 0.03  & 0.45  &  0.330  \\
Fe 4383   &  -0.23   & 0.42	 & 1.04 & 0.07  & 0.83  &  0.838 \\
Ca 4455   &  -0.12   & 0.09      & 1.21 & 0.05  & 0.36  &  0.191  \\
Fe 4531   &  0.94    & 0.23	 & 0.84 & 0.05  & 0.48  &  0.428  \\
C$_2$ 4668&  0.67    & 0.32	 & 0.94 & 0.05  & 0.76  &  0.896 \\
H$\beta$  &  0.125    & 0.073    & 1.10 & 0.02  & 0.30  &  0.217 \\
Fe 5015   &  0.62    & 0.27	 & 1.02 & 0.04  & 0.63  &  0.495  \\
Mg b      &  0.007   & 0.094     & 1.06 & 0.03  & 0.17  &  0.218 \\
Fe 5270   &  0.16    & 0.15	 & 1.00 & 0.05  & 0.35  &  0.280  \\
Fe 5335   &  0.28    & 0.15	 & 1.12 & 0.05  & 0.34  &  0.307  \\
Fe 5406   &  0.29    & 0.10      & 1.02 & 0.05  & 0.21  &  0.244  \\
Mg$_1$    &  -0.044  & 0.003     & 0.95 & 0.02  & 0.009 &  0.0090  \\
Mg$_2$    &  -0.050  & 0.004     & 0.98 & 0.01  & 0.010 &  0.0102  \\
H$\gamma$a&  -0.32   & 0.19	 & 0.95 & 0.02  & 0.86  &  0.753  \\
H$\gamma$f&   0.000   & 0.066    & 1.15 & 0.02  & 0.39  &  0.287  \\
Ca 1       &  0.157  &   -  &	-   &	-       & 0.194 &  0.812  \\
Ca 2       &  0.454  &   -  &	-   &	-       & 0.260 &  0.779  \\
Ca 3       &  0.512  &   -  &	-   &	-       & 0.413 &  0.785  \\
CaAZ1	  & -  &   -  &	-   &	-       & -            &   0.365  \\
CaAZ2	  & -  &   -  &	-   &	-       & -            &   0.381  \\
CaAZ3	  & -  &   -  &	-   &	-       & -            &   0.369  \\
CaAD1	  & -  &   -  &	-   &	-       & -            &   0.429  \\
CaAD2	  & -  &   -  &	-   &	-       & -            &   0.386  \\
CaTP1     & -  &   -  &	-   &	-       & -            &   0.389  \\
CaTP2     & -  &   -  &	-   &	-       & -            &   0.428  \\
CaTP3     & -  &   -  &	-   &	-       & -            &   0.387  \\
\hline 
\end{tabular}
\end{center}
}
\end{table}

Having established the accuracy of our {\sl average} fiber measurement, 
we then investigated systematic fiber-to-fiber differences using  
out-of-focus stars. In principle differences in spectral response can occur
because each of the fibers sees a different area of the primary
mirror, but moving the out-of-focus stars slightly in the focal
plane showed that this effect was negligible. The out of focus stars 
were reduced, and for each fiber absorption line indices 
were determined.  Taking all the fibers for which the integrated flux 
was larger than 30\%  of the flux of the brightest fiber we determined
the RMS scatter in the line indices for each star, after applying a 
3$\sigma$ rejection algorithm. These values are given  in the last 
column in Table 3.  The RMS scatter that we found
only slightly depends on the signal-to-noise ratio in the stars, 
showing that these
fiber-to-fiber differences are due to some systematic effect. Careful
inspection of the data shows that  fringing is important in the data,
especially in the red, but also somewhat in the blue, and that incorrect
flatfielding can affect the values of the indices severely. We therefore think
that residual fringing (which is  dependent on the position of the telescope)
might be responsible for our small residual variations, although it is
not excluded that variations in fiber stresses leading to small variations 
in the continuum could also be responsible. Structures smaller than this
level in our maps  should probably not be believed.
The RMS fiber to fiber differences are on the order of 0.3 ${\rm \AA}$ and
0.010 for the indices expressed in magnitudes. This means that very small
gradients are hard to determine, but in general this accuracy is good enough
to obtain results of the quality currently produced in the 
literature (see the comparison with Paper II in the next section).

As a last step we determine maps for each absorption line index, as
well as for the integrated continuum, radial velocity, velocity dispersion
and higher order velocity moments. The kinematics will be described in 
a following article (Prada \etal 1999, in preparation). Maps were reconstructed
using triangular interpolation and regridded to frames with
a pixelsize of 0.48''. Since the data are undersampled the maps will contain
some small amount of aliasing. However, the comparison of the reconstructed
intensity maps with HST data (see Section 4.1) shows that this is minor 
and hard to detect.
We see however that in the three galaxies the position of the galaxy center
in the red is somewhat different from in the blue. This is due to differential
refraction (see for an example  Arribas \etal 1997). The differential
refraction within the red or blue band however was so small compared to the 
spatial resolution that we didn't correct for it. 
To give an idea about the final obtained resolution Arribas \etal (1997)
looked at reconstructed images of in focus stars, and found them to have 
FWHM values of 1.2''. For the data in this paper the analysis of Section
4.1 shows that similar values have been obtained here.

In Fig.~3 we show all line index maps of the three galaxies. To make
the interesting features visible we have for each map subtracted 
its median value, and then divided all pixels by the standard deviation
of the map. Many maps only show noise, indicating that the line
index everywhere in this central region of the galaxy is the same.
However, also some prominent features are seen. These will be discussed
in Sections 4 and 5.

\onecolumn
\begin{figure}
\mbox{\epsfxsize=16cm  \epsfbox{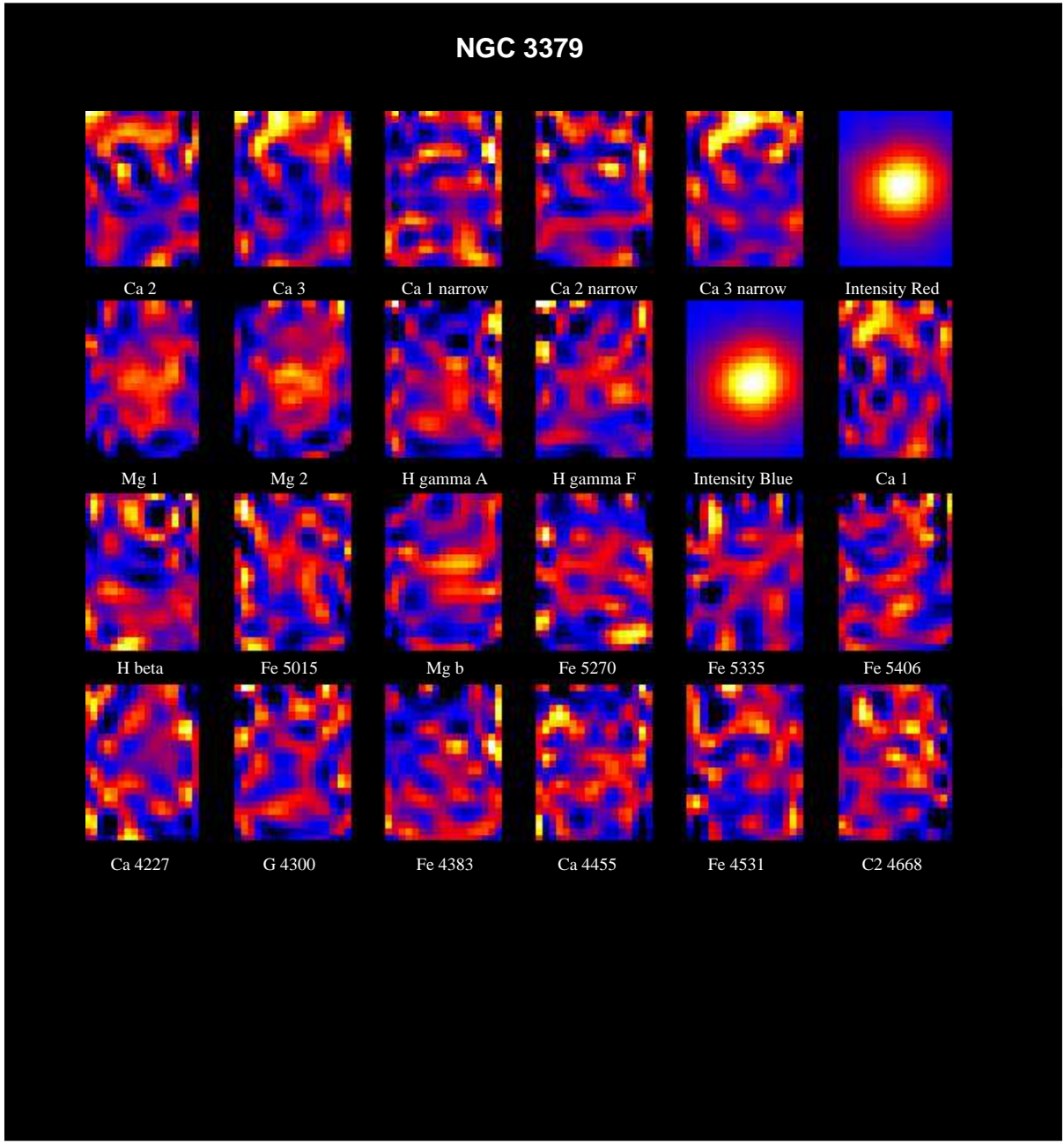}}
\caption{a: Maps of the most important line indices in blue and red for NGC
3379.
The maps have been scaled in such a way that pixels that are more
than 2 $\sigma$ fainter than the mean value of a map are black, while 
those more than 2 $\sigma$ brighter are white. If there is very little
structure in a map, it will look noisier than in a case where the dynamic
range in the map is large. The size of the map is 8.2$''$ $\times$ 11.0$''$}
\end{figure}
\addtocounter{figure}{-1}
\begin{figure}
\mbox{\epsfxsize=16cm  \epsfbox{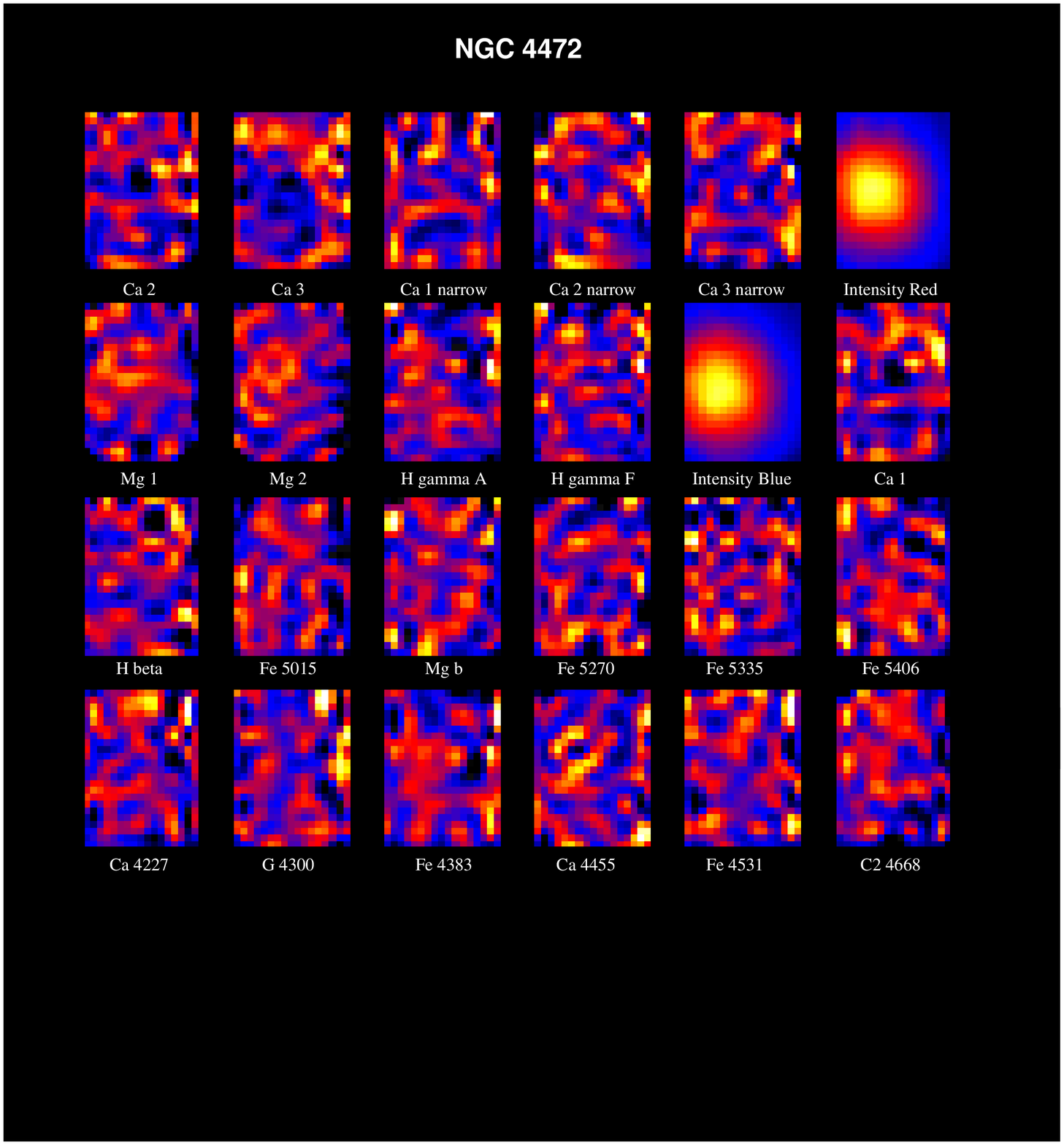}}
\caption{b: Maps of the most important line indices in blue and red for NGC
4472.}
\end{figure}
\addtocounter{figure}{-1}
\begin{figure}
\mbox{\epsfxsize=16cm  \epsfbox{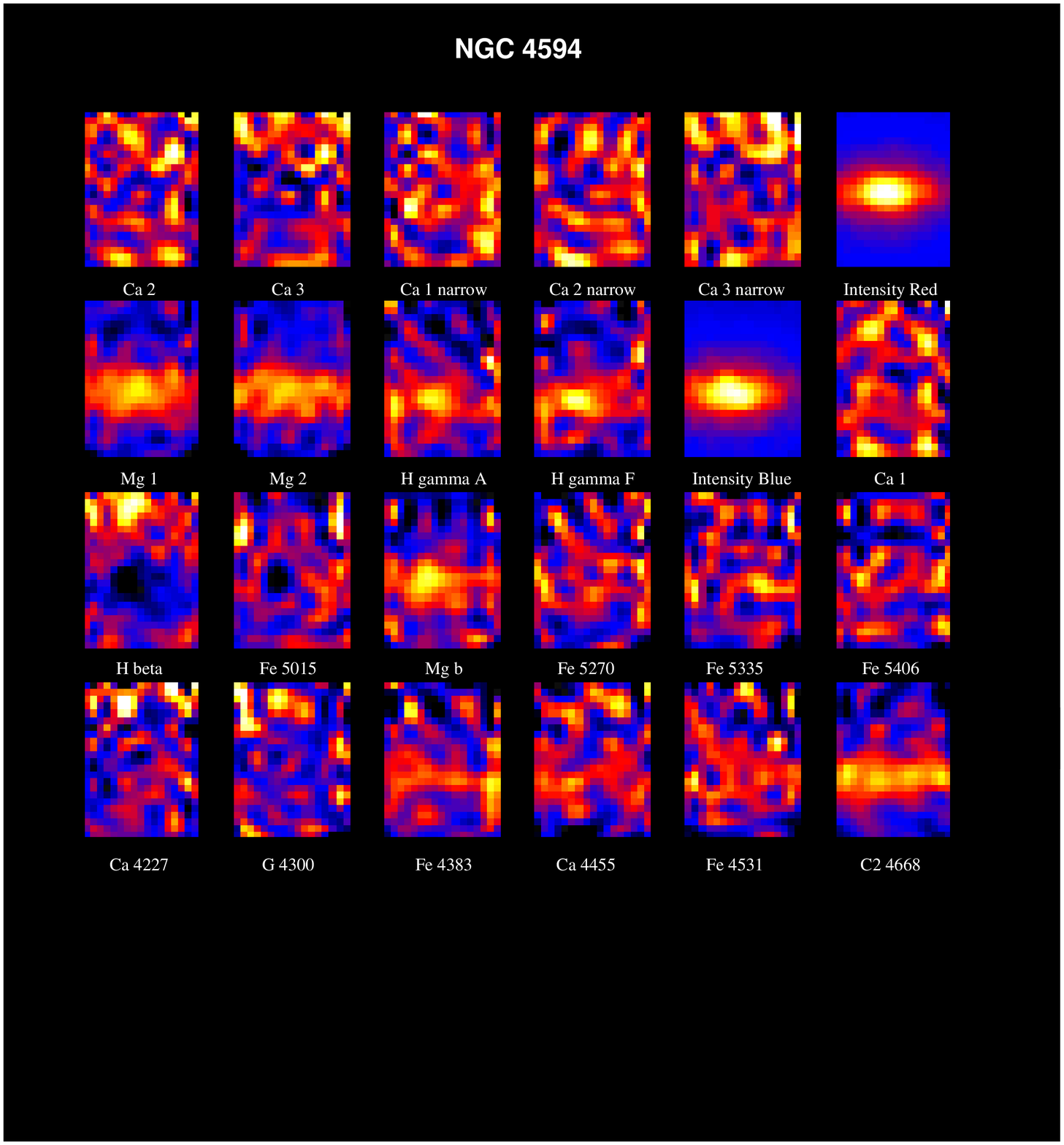}}
\caption{c: Maps of the most important line indices in blue and red for NGC
4594.}
\end{figure}
\twocolumn

\section{Comparison with the Literature}

\subsection{Continuum images}

The first test we applied was to reconstruct the continuum images
in red and blue, fit ellipses to these images, and compare them with
recent HST archive images. The spectra were collapsed in the red
between 8427 and 8761 ${\rm \AA}$, and in the blue between 4236 and 5643
${\rm \AA}$. On the reconstructed images (see Fig.~3) ellipses were fit 
using Galphot (see J\o rgensen et al. 1992).  In Fig. 4 we show the blue surface 
brightness profiles, normalized in the center, of the three galaxies, and the 
profiles of the ellipticity (1-$b\over{a}$), and the major axis position angle. 
Also shown are profiles determined on recent HST-WFPC2 images in 
V (F555W), convolved to a seeing of 1.2'' (FWHM). Plotted on
the abscissa is $\sqrt{{\rm major} \times {\rm minor~ axis~ radius}}$, 
while the surface brightness
profiles are scaled by an arbitrary number.  The reconstructed images,
together with the convolved HST frames, are shown in Figure~5.

\begin{figure}
\begin{center}
\mbox{\epsfxsize=6cm  \epsfbox{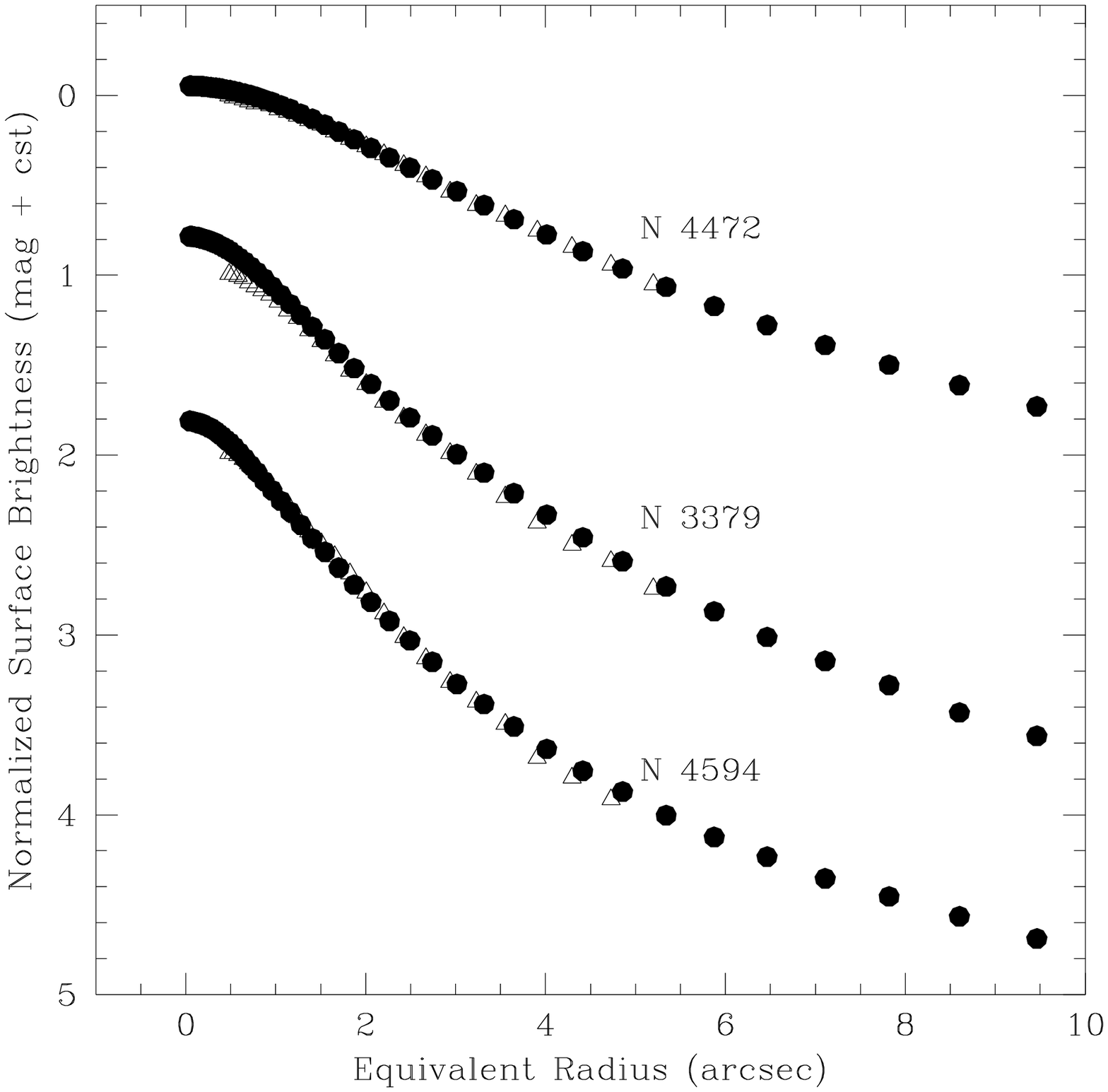}}
\mbox{\epsfxsize=6cm  \epsfbox{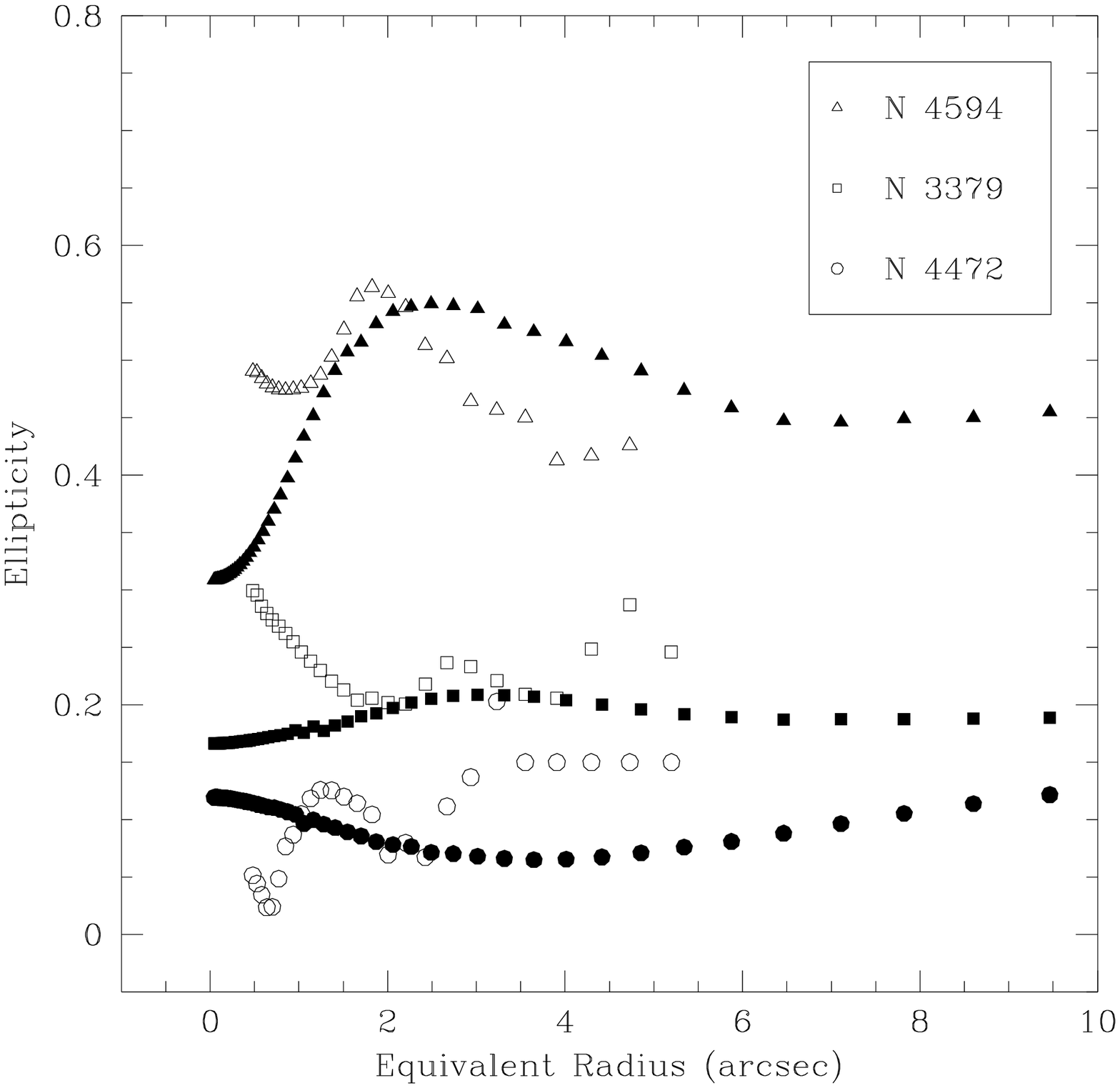}}
\mbox{\epsfxsize=6cm  \epsfbox{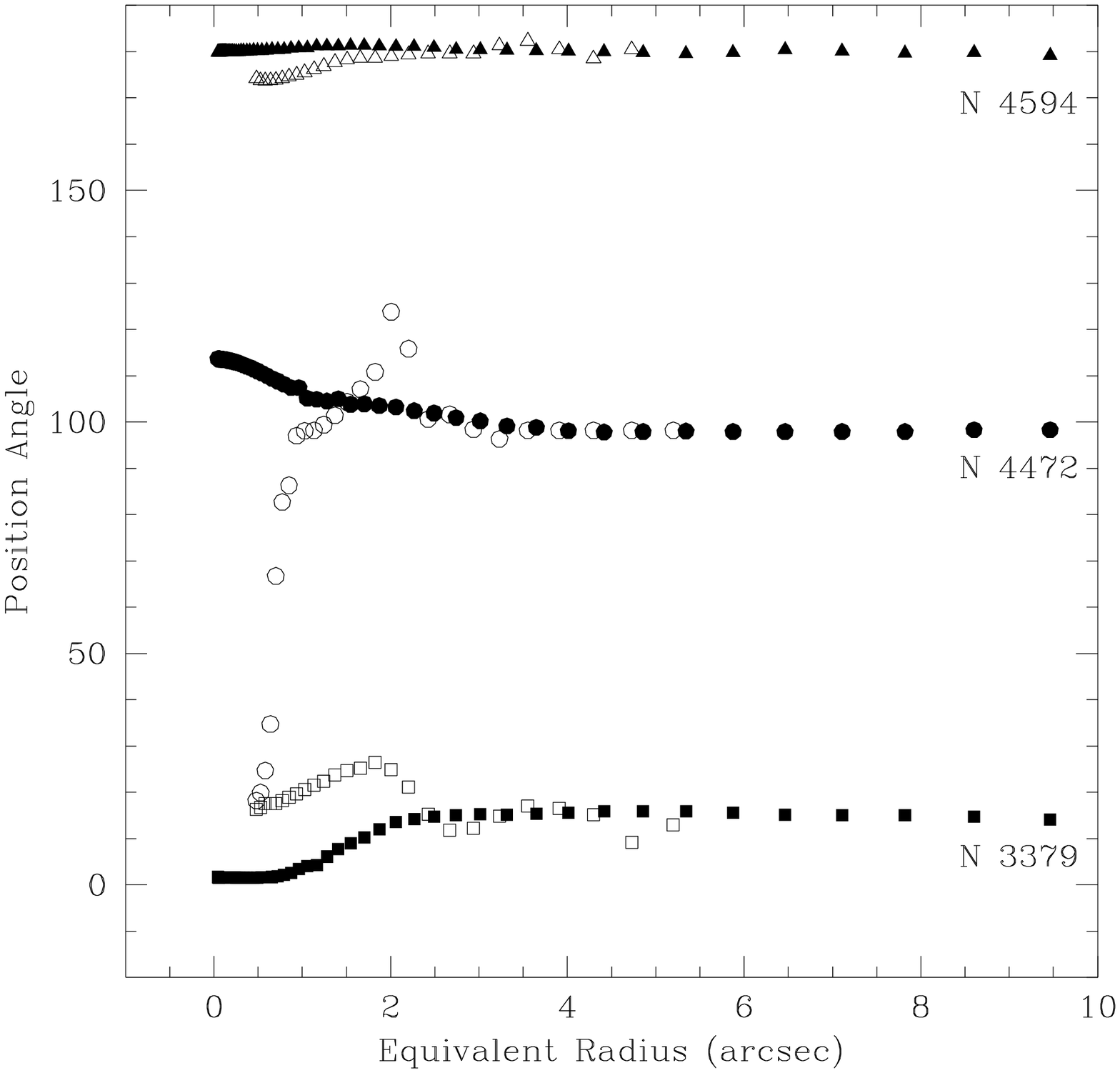}}
\caption{{\bf a:} Normalized blue surface brightness profiles compared 
to HST profiles in V, convolved to a seeing of 1.2'' FWHM. Filled symbols
indicate HST data; open symbols data from this paper. 
{\bf b:} Ellipticity profiles.
For clarity 0.1 has been added to the ellipticities
of NGC~3379. {\bf c:} Position Angle profiles. Arbitrary offsets have been
applied
to the HST position angle zeropoints. }
\end{center}
\end{figure}

\begin{figure}
\begin{center}
\mbox{\epsfxsize=12cm  \epsfbox{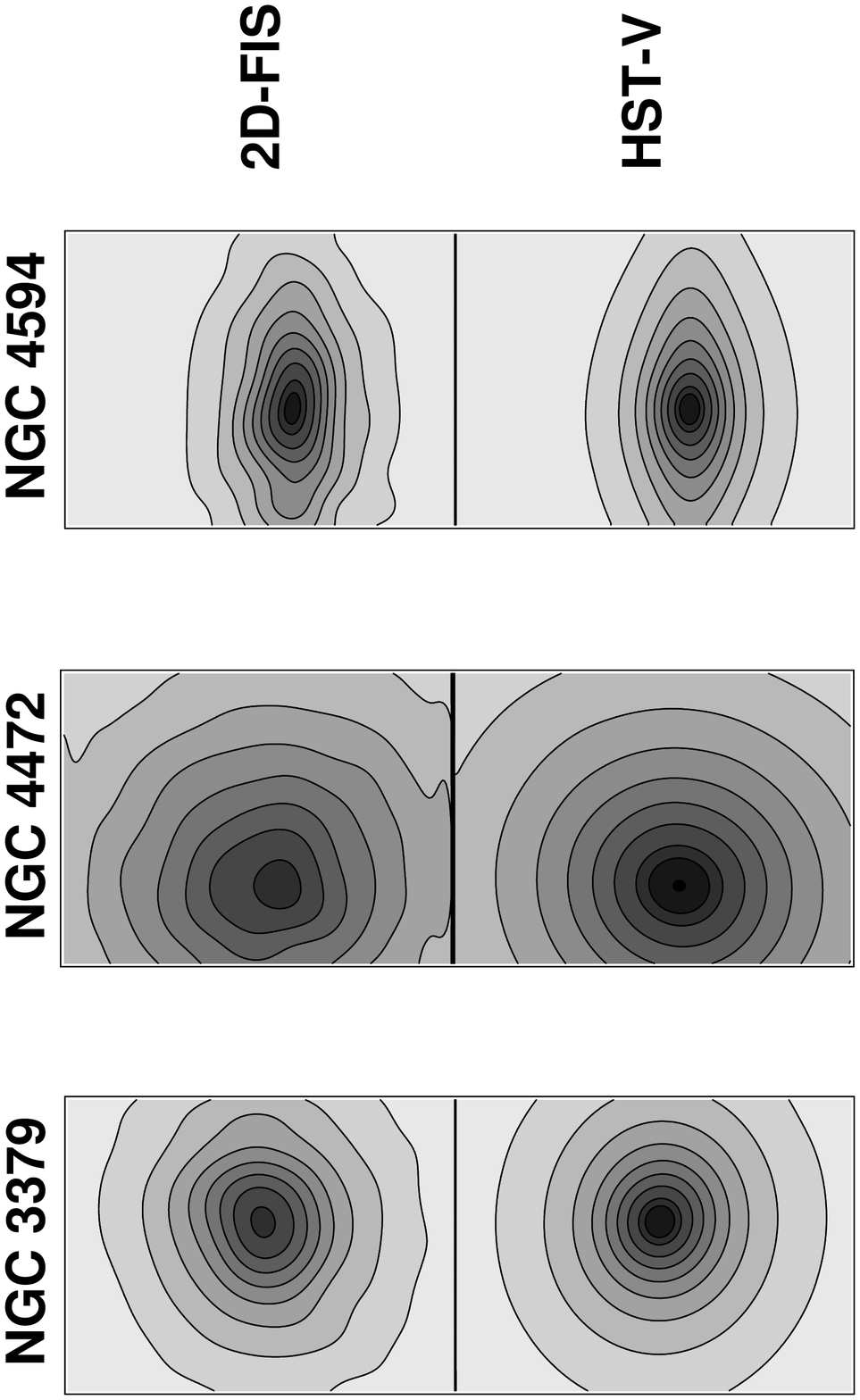}}
\caption{Comparison of our reconstructed continuum images in the blue
with HST V-band images, convolved to a seeing of 1.2''. The intensities
of both images have been scaled in the outer parts. Isophote levels
are 0.2,0.3,0.4,0.5,0.6,0.7,0.8,0.9,0.95 and 1.00 times the maximum of
the HST images. The size of the images is 8.2$''$ $\times$ 11.0$''$, the same
as Fig.~3.}
\end{center}
\end{figure}

The agreement between the continuum images and the convolved HST V-band 
images is good. Their slope can be seen to be slightly shallower,
which is what one expect from colour gradients, since the effective
wavelength of our images is somewhat bluer than 5550 ${\rm \AA}$.
The fact that the ellipticity profiles beyond 3'' start diverging is
due to the fact that here not the whole ellipse in included in our 
2D-FIS images. Agreement in the red is not as good as in the blue,
but still good. We conclude that the good agreement between our photometry
and HST shows that stray light is not a problem in the analysis
of our spectra.

\subsection{Comparison with long-slit spectra}

One of the main reasons why we are studying these 3 galaxies in such detail,
is that accurate line strength profiles are available for most
of the Lick indices from Paper II. In this paper the major axis of NGC~3379 
and NGC~4472, and the  minor axis
of NGC~4594 were observed with the same spectrograph, ISIS, on the WHT, 
but using its regular long-slit mode. The
seeing of those observations was not as good as for this  paper - their
effective seeing was about 2.5'' - 3''. The spectra  of Paper II
were calibrated onto the Lick system and  corrected for velocity dispersion
in the same way as was done for the data in the current paper. For the
comparison we turned  the images of NGC 4472, NGC 3379 and NGC4594 by resp.
-10, 75 and 0$^{\rm o}$, and then  extracted the line strength 
profiles in a column of
width 1'' through the center. The comparison is given in Fig.~6. In
general the comparison is good, although some comments have to be made. The
errorbars in the upper  right corner take into the account the fiber-to-fiber
errors and the error made by converting to the Lick system. One can see that
most of the discrepancies can be explained by this error. One also can see
that the seeing for our observations was much better - this is clear from
e.g. H$\beta$, \mg2  and Mg$b$ for the Sombrero galaxy, where the inner  disk
is now much better resolved. H$_{\beta}$ and Fe 5015 in the center of the 
Sombrero galaxy are lower than elsewhere, because of contamination 
by emission; in case of the Fe 5015 line by the [OIII] line at 
5007${\rm \AA}$.
Agreement for well-studied indices,
like \fe\ and \mg2 is excellent. Apart from these differences there appear to be
some small global offsets, in Ca 4227, and maybe somewhat in the G-band.
Our Ca 4227 is now in better agreement with the models of e.g. Worthey
\etal (1994) than the data of Paper II, but the line strength is still 
not as large as one would expect (see Paper II).

Sometimes small 'jumps' are seen in our index-profiles, like
e.g. at -5'' for the Sombrero galaxy, due to the problems described in 
Chapter 3. Apart from these, our index profiles are as good
as most currently available profiles in the literature.

\onecolumn
\begin{figure}
\mbox{\epsfxsize=16cm  \epsfbox{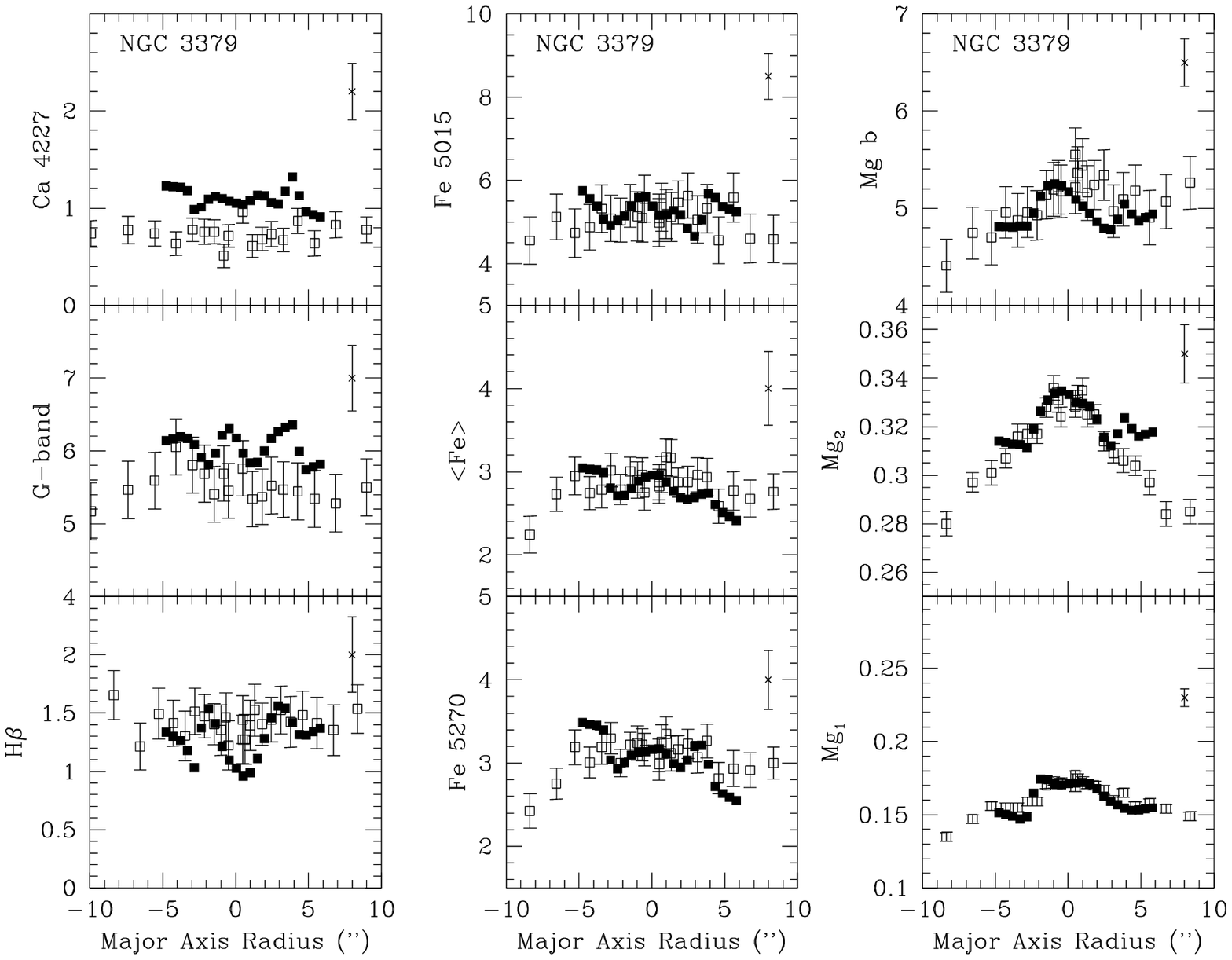}}
\caption{Comparison of our blue indices (filled symbols) 
with the long-slit spectra of Paper II (open symbols).}
\end{figure}
\addtocounter{figure}{-1}
\begin{figure}
\mbox{\epsfxsize=16cm  \epsfbox{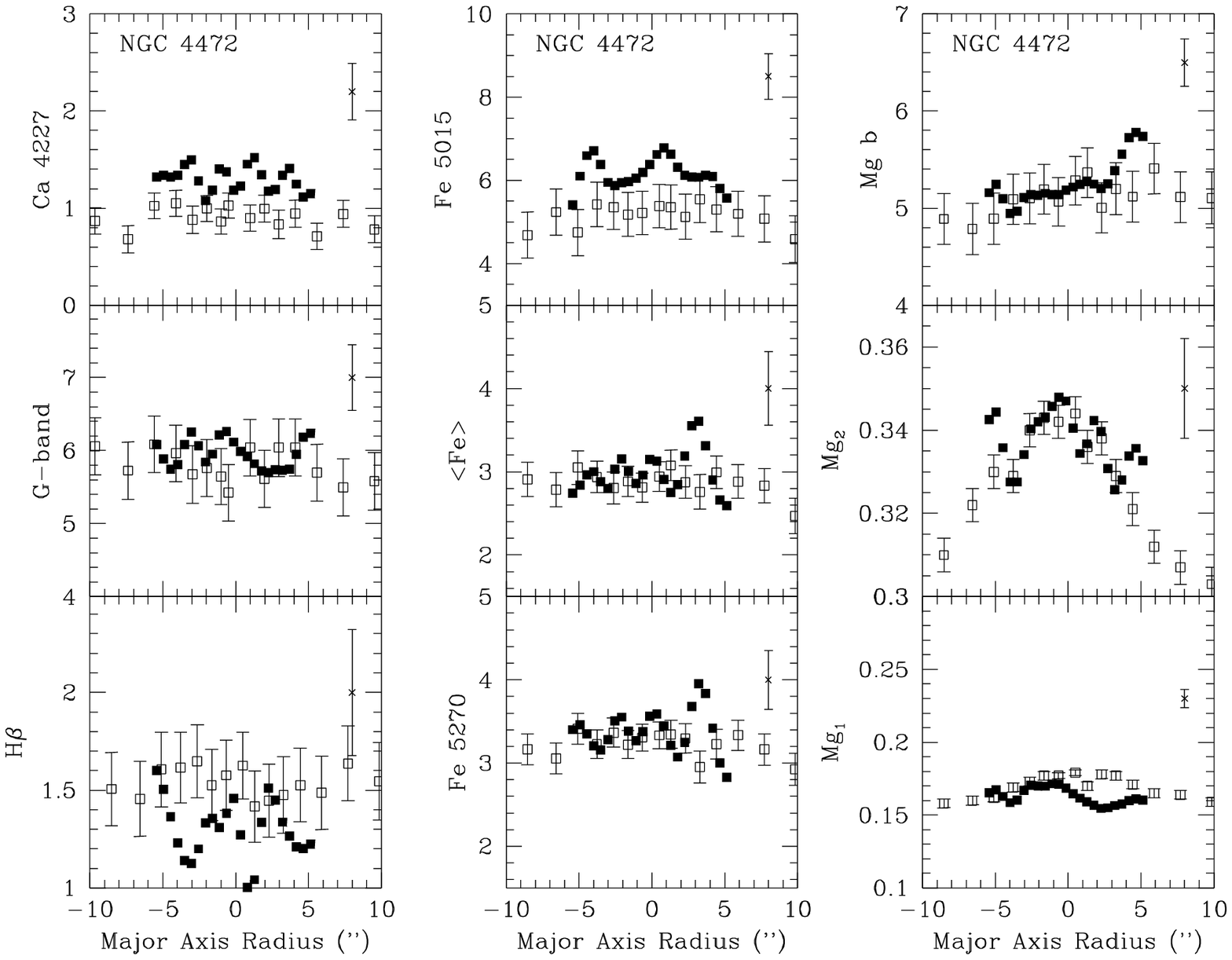}}
\caption{Comparison of our blue indices (filled symbols) 
with the long-slit spectra of Paper II (open symbols).}
\end{figure}
\addtocounter{figure}{-1}
\begin{figure}
\mbox{\epsfxsize=16cm  \epsfbox{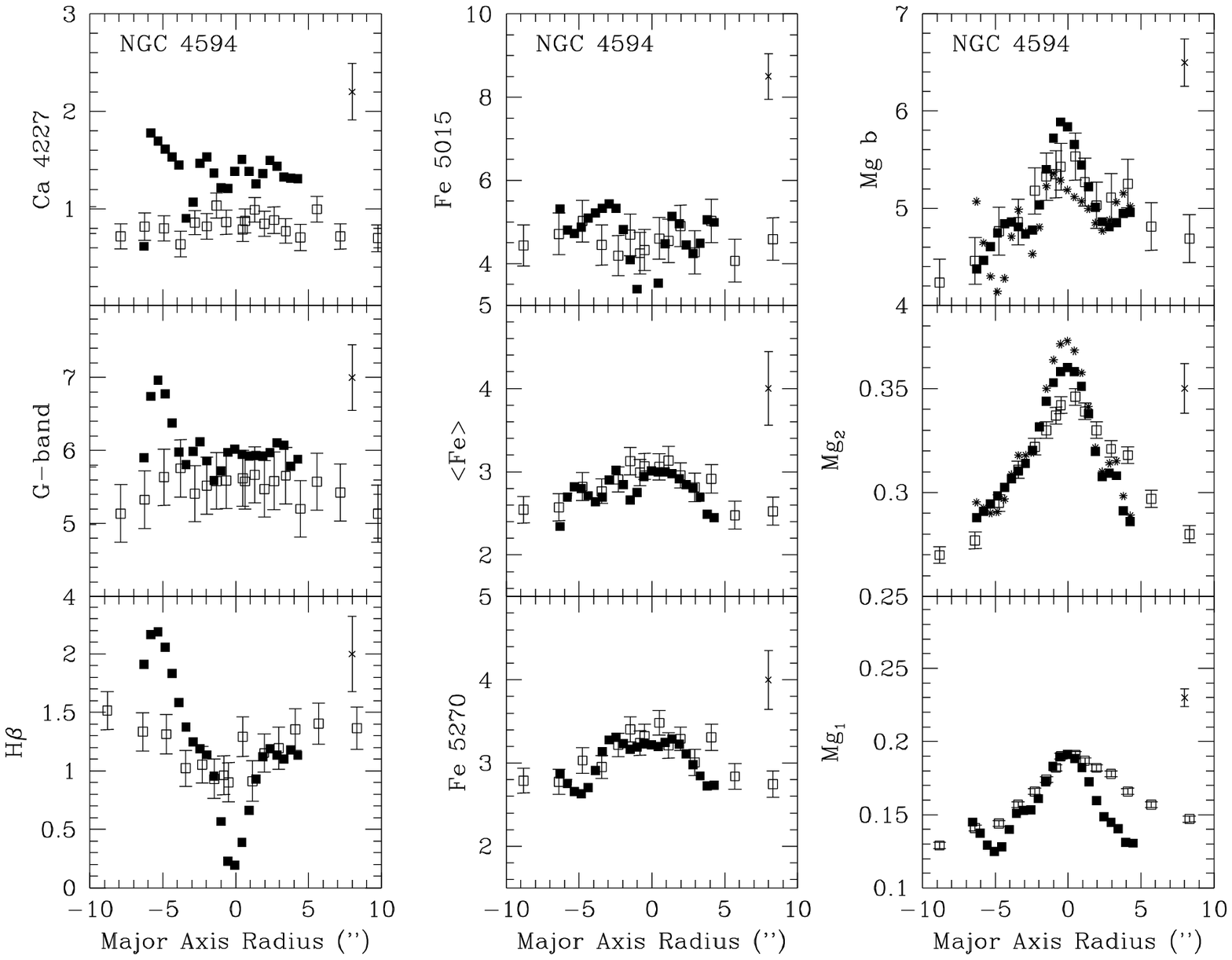}}
\caption{Comparison of our blue indices (filled symbols) 
with the long-slit spectra of Paper II (open symbols). 
The profiles corrected for [NII] emission are indicated by asteriscs. }
\end{figure}
\twocolumn

\subsection{Comparison with TIGER}

The only observations of a similar kind available in the literature
are those taken with TIGER, an instrument
built in Lyon (Bacon \etal 1995). Tiger uses multilenslets, and compared
to our current observations has a better sampling, but a much smaller
wavelength range. Two detailed two-dimensional studies of the 
stars have been published (Emsellem \etal (1996) \& Bacon \etal (1994)),
of which the first deals with the Sombrero galaxy. 
The agreement with
Emsellem's figures 8 and 9 is good. In our study the same kinematic
features are found. Although we will make a more detailed
comparison in a next paper (Prada et al., in 
preparation), we show here that the radial velocitiy and the velocity dispersion 
along
the major axis agree within 10 km/s (Fig.~7). 
In Emsellem \etal (1996) a comparison
is made with Kormendy \etal (1988), which also is satisfactory.

\begin{figure}
\mbox{\epsfxsize=8cm  \epsfbox{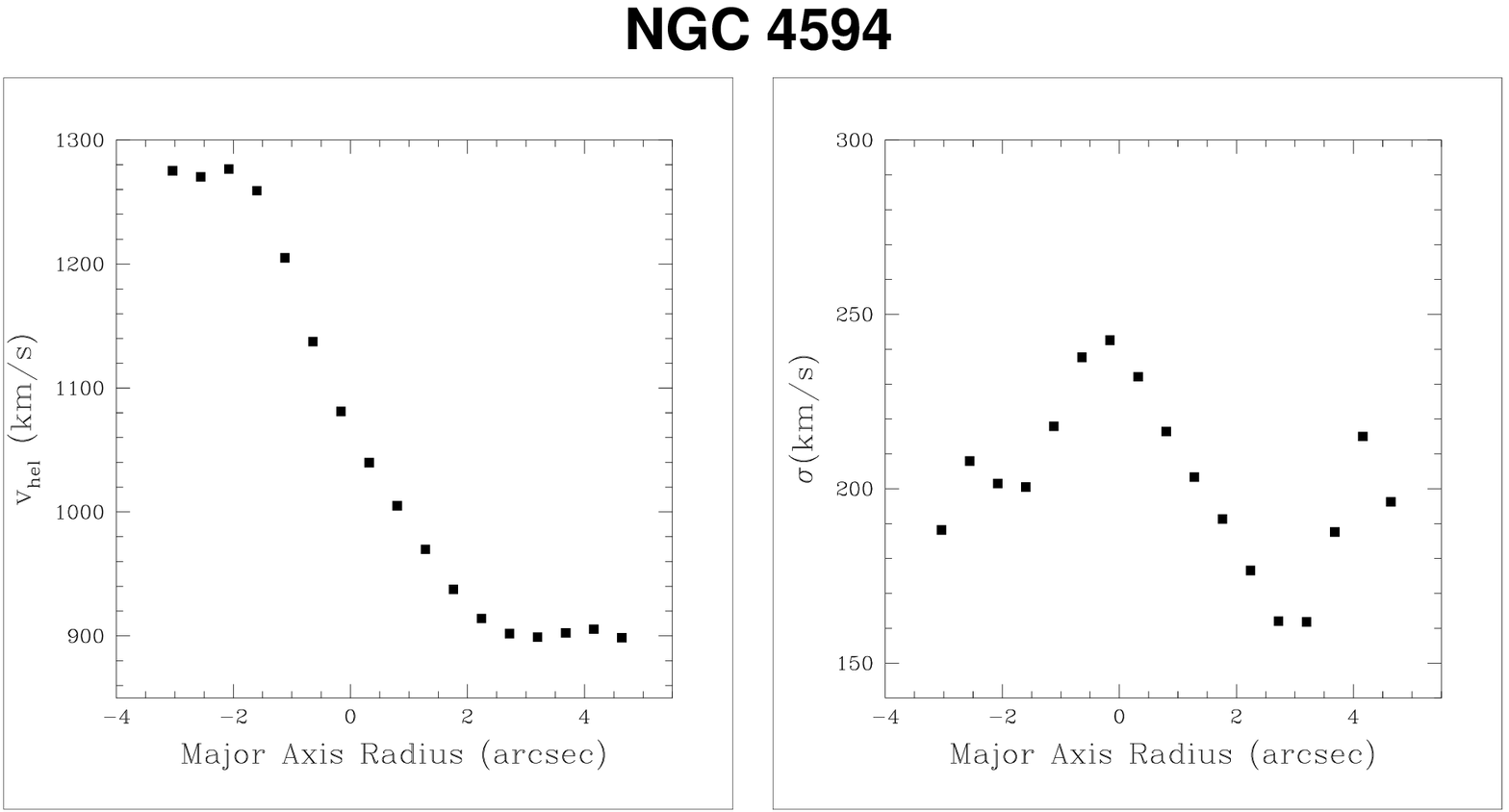}}
\caption{Major axis cuts of radial velocity and velocity dispersion of the 
Sombrero galaxy, from the blue datacubes. See Emsellem et al. (1996) for
a comparison with TIGER data.}
\end{figure}

To compare our Mg$b$ image with Emsellem \etal (1996)'s image, we first
corrected it for emission lines. As explained in Goudfrooij \& Emsellem
(1996) one of the wings of this line includes an emission line
of [NI] at 5199 ${\rm \AA}$, which has the effect of increasing the continuum
and so also the feature's strength. This emission is relatively strong
in the Sombrero (Ho et al. 1997).
We used the fact that this feature
is very sharp, as compared to the wide stellar features, and removed
it by interpolating across it, as one does for bad pixels. The results
of this operation can be seen for Mg~b and \mg2 (which is also slightly
affected) in Fig.~ 6. The agreement between our Mg$b$ map and 
Fig.~18 of Emsellem \etal\ is good, reproducing in both cases
the inner disk, which is strong in Mg$b$, and the peak in the center.

Since the strength of the emission line drops off quickly away from 
the center, it is possible to obtain an emission line spectrum by taking an 
average spectrum at about 2-3$''$ from the center, scaling the continuum, 
adjusting the velocity dispersion, and 
subtracting it from the central spectrum.  
Some emission was also found in the center of 
NGC~3379, but much fainter, and not strong enough to detect the 
[NI] 5199 ${\rm \AA}$ line, and less extended.


\section{Results}

\subsection{Radially Averaged Profiles}

In this section we present radially averaged profiles of the 
3 galaxies. They have been averaged on the ellipses calculated
on the continuum images. We see in Fig.~8 that the profiles of the three 
galaxies generally lie on top of each other, showing that the three
galaxies have similar stellar populations (see also Paper II). 
An exception is the inner region of the Sombrero galaxy,
not only because of the emission lines, but also because we see large
colour gradients in lines like e.g. H$\gamma$ and \mg2.
In agreement with long slit studies most indices have non-zero radial
gradients. These gradients are given in Table~4.
In the red the Ca T indices seem to increase somewhat going outward,
consistent between the three of them.

\onecolumn
\begin{figure}
\mbox{\epsfxsize=16cm  \epsfbox{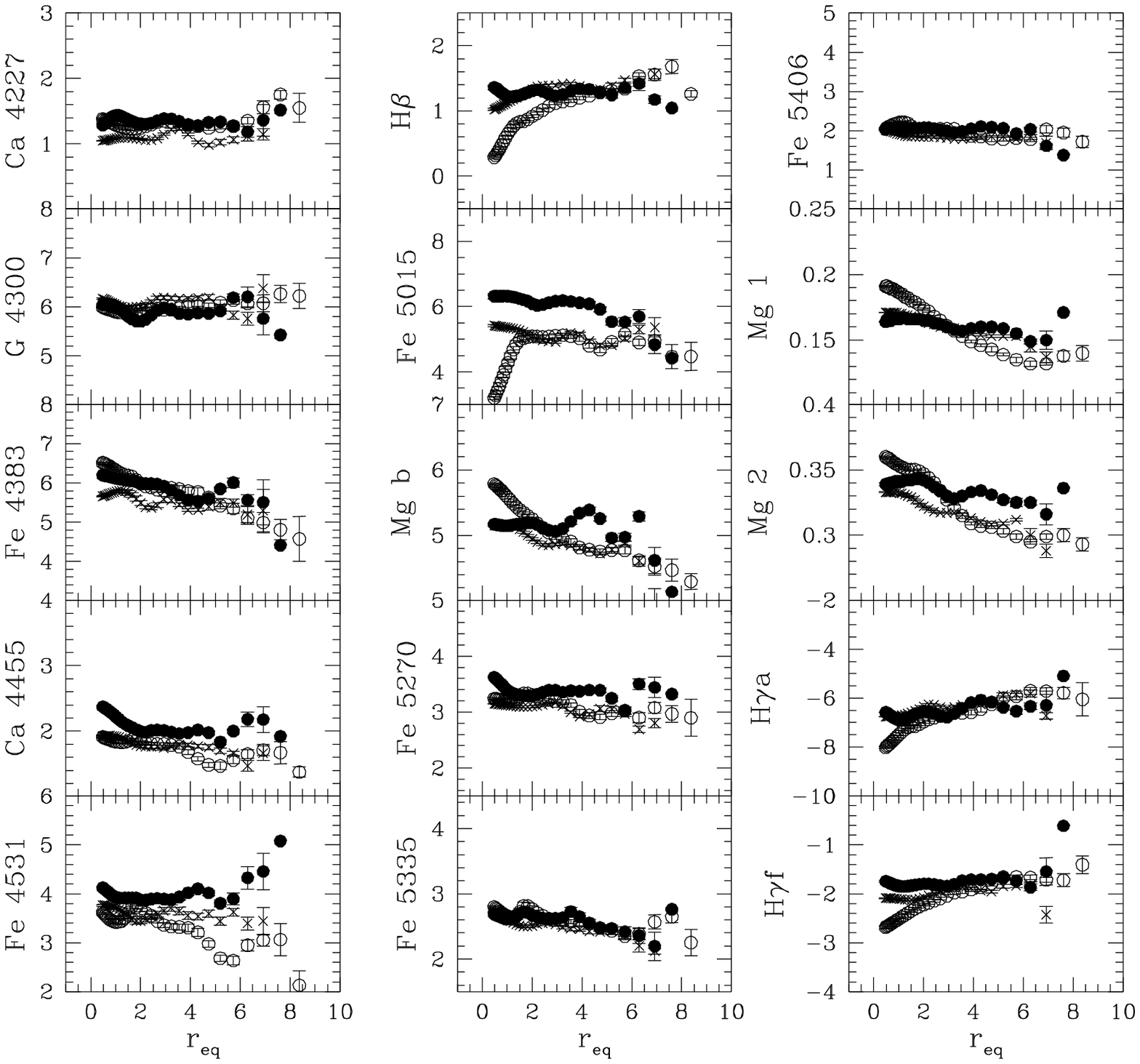}}
\caption{a: Radially averaged line indices in the blue. See Fig.~8b for the
definition of the symbols.}
\end{figure}
\addtocounter{figure}{-1}
\begin{figure}
\mbox{\epsfxsize=16cm  \epsfbox{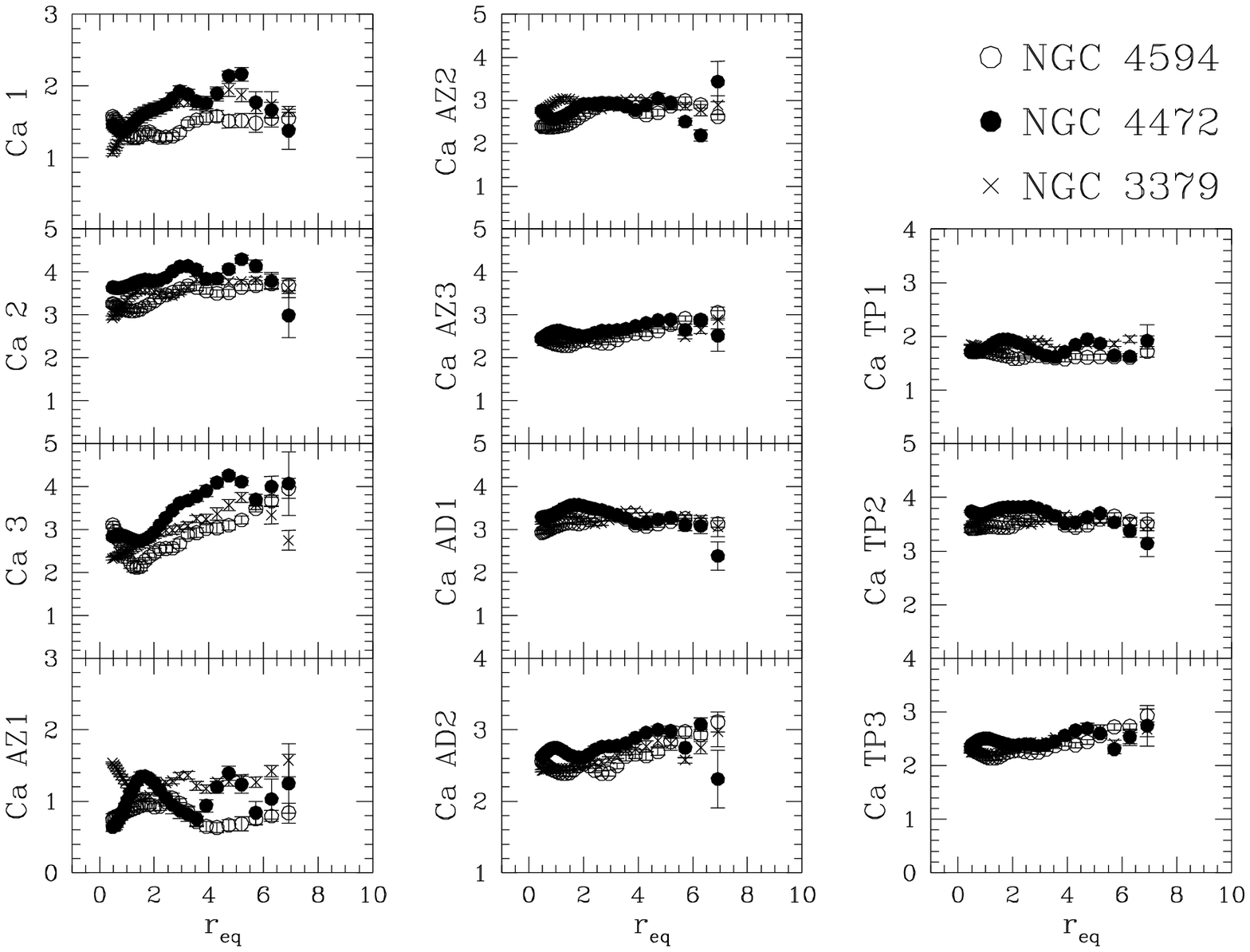}}
\caption{b: Radially averaged line indices in the red.}
\end{figure}
\twocolumn

\begin{table}
{\small
\begin{center}
\caption{Line strength gradients. Results are given for simple
linear fits of the line strengths as a function of the logarithm of
the radius, using the radially average line strength profiles. Tabulated
are the fitted line strength at r$_{eq}$ = 1$''$ (a) and the gradient per
dex (b). }
\begin{tabular}{lcccccc}
\hline
\hline
Band & \multicolumn{2}{c}{NGC 3379} & \multicolumn{2}{c}{NGC 4472}  &
\multicolumn{2}{c}{NGC 4594} \\
   &  a & b & a & b & a & b \\
\hline
Ca 4227   &  	 1.078  &     0.036	&  1.359  &    -0.028	&    1.303   &   0.067     \\
G 4300    &  	 6.086  &    -0.020	&  5.964  &    -0.170	&    5.932   &	  0.158    \\
Fe 4383   &  	 5.680  &    -0.410	&  6.114  &    -0.739	&    6.308   &	 -1.208    \\
Ca 4455   &  	 1.887  &    -0.277	&  2.208  &    -0.336	&    1.857   &	 -0.275    \\
Fe 4531   &  	 3.680  &    -0.251	&  3.965  &	0.204	&    3.531   &	 -0.558    \\
H$\beta$  &  	 1.150  &     0.421	&   1.286 &    -0.040	&     0.613  &	   1.064   \\
Fe 5015   &  	 5.300  &    -0.394	&  6.295  &    -0.894	&    5.347   &	 -0.639    \\
Mg b      &  	 5.110  &    -0.600	&  5.175  &    -0.228	&    5.543   &	 -1.133    \\
Fe 5270   &  	 3.126  &    -0.215	&  3.444  &    -0.193	&    3.231   &	 -0.243    \\
Fe 5335   &  	 2.624  &    -0.364	&  2.660  &    -0.170	&    2.715   &	 -0.252    \\
Fe 5406   &  	 1.923  &    -0.196	&  2.065  &    -0.199	&    2.109   &	 -0.275    \\
Mg$_1$    &  	 0.169  &    -0.023	&  0.165  &    -0.009	&    0.184   &	 -0.055    \\
Mg$_2$    &  	 0.330  &    -0.031	&  0.340  &    -0.014	&    0.352   &	 -0.060    \\
H$\gamma$a&  	-6.640  &     0.672	& -6.734  &	0.630	&   -7.508   &	  1.924    \\
H$\gamma$f&  	-2.078  &     0.172	& -1.815  &	0.254	&   -2.475   &	  0.950    \\
Ca 1      &  	 1.405  &     0.634	&  1.506  &	0.477	&    1.412   &	  0.053    \\
Ca 2      &  	 3.287  &     0.710	&  3.725  &	0.276	&    3.221   &	  0.537    \\
Ca 3      &  	 2.559  &     1.076	&  2.907  &	1.310	&    2.592   &	  0.644    \\
CaAZ1	  &  	 1.294  &    -0.056	&  0.929  &	0.314	&    0.846   &	 -0.046    \\
CaAZ2	  &  	 2.893  &     0.050	&  2.699  &	0.244	&    2.470   &	  0.527    \\
CaAZ3	  &  	 2.418  &     0.329	&  2.562  &	0.221	&    2.363   &	  0.393    \\
CaAD1	  &  	 3.247  &     0.037	&  3.397  &    -0.279	&    3.070   &	  0.234    \\
CaAD2	  &  	 2.467  &     0.363	&  2.695  &	0.153	&    2.461   &	  0.350    \\
CaTP1     &  	 1.812  &     0.060	&  1.778  &	0.039	&    1.694   &	 -0.115    \\
CaTP2     &  	 3.589  &     0.004	&  3.740  &    -0.209	&    3.453   &	  0.162    \\
CaTP3     &  	 2.265  &     0.385	&  2.438  &	0.111	&    2.223   &	  0.374    \\
\hline 
\end{tabular}
\end{center}
}
\end{table}

\subsection{Two-Dimensional Features in the line strength maps}

One of the main advantages of IFS is that one can rather easily make
absorption line maps, and use them  to derive conclusions
about galaxy formation. In this paper we have found that the inner 
disk of the Sombrero is very well visible in the Mg maps (Mg$_1$, Mg$_2$ and Mgb),
but is almost invisible in maps of Fe and Ca indices. This is mainly due 
to the fact the relative noise in the Fe and Ca maps is much larger than
in the Mg features. To show this, we show in Fig.~9 a diagram of Mg$_2$ vs.
$<$Fe$>$ that includes the nuclei of some well-observed ellipticals and
spirals, some models of Vazdekis et al. (1996), and the radial profile
of the Sombrero galaxy (see Section 5.1). The figure has been taken from
Peletier (1999). One can see that the slope of the Sombrero is very similar
to that of the models, indicating that in the Sombrero Mg/Fe remains 
constant, although [Mg/Fe] is larger than solar.  
This results appears surprising at first view. A much used explanation
of overabundant Mg/Fe ratios is that the enrichment of Mg is mainly due to
massive stars producing SN type II, while Fe is mainly produced in SN type
Ia (see e.g. Worthey et al 1992). 
Since the latter originate in binaries, there is a time delay of a few
times 10$^8$ year between star formation and release of the Fe-peak elements
in those SNe type Ia. Elliptical galaxies originally were thought to form
their stars on very short timescales, as opposed to disks, and for that reason
ellipticals would have enhanced Mg/Fe, while Mg/Fe in disks would be around solar.
In an excellent review
paper Worthey (1998), summarizing the current observational status of Mg/Fe
in external galaxies, shows that this explanation is probably incorrect.
Although giant ellipticals clearly have supersolar Mg/Fe ratios, this does
not seem to be the case for fainter ellipticals (e.g. Davies et al. 1993),
whose Mg/Fe are close to what is predicted by models with solar abundance 
ratios. Also, for spirals the situation is not so clear any more. Jablonka
et al. (1996) find that their most luminous bulges have enhanced [Mg/Fe],
but fainter ones do not. Sil'chenko (1993) finds that [Mg/Fe] is solar
in most bulges of other spiral galaxies, although her sample contains very
few massive galaxies. Based on these and other results Worthey concludes 
that the data seems to indicate that
[Mg/Fe] is near zero in most galaxies of all types with velocity dispersion
less than about 225 km/s, and that galaxies with velocity dispersions above 
that values are progressively more Mg-enhanced. Peletier (1999) slightly
refines this for spirals, for which the potential is not determined 
only by random motions, replacing the words 'velocity dispersion' by
kinetic energy, or escape velocity.

\begin{figure}
\begin{center}
\mbox{\epsfxsize=8cm  \epsfbox{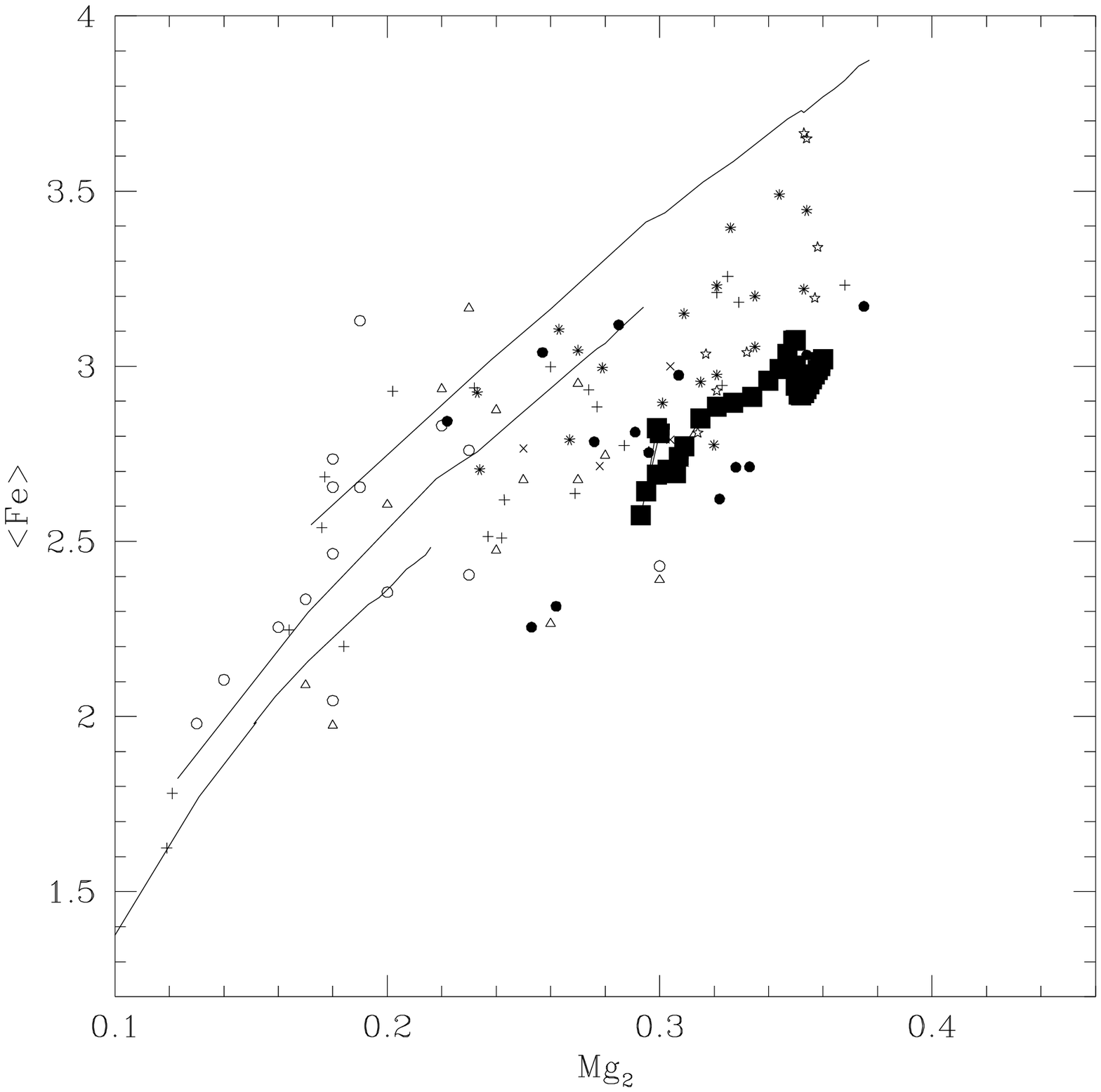}}
\caption{Literature compilation of Mg$_2$ vs. $\langle$Fe$\rangle$. Plotted
are central values. The meaning of the symbols is as follows:
open circles are bulges of spiral galaxies (morphological type 3-5) from
Jablonka et al. 1996, open triangles bulges of type 0-2 from the same
paper, crosses galaxies from Carollo et al. (1993), asteriscs from Fisher
et al. (1996), pentagons from Davies et al. (1993), filled circles from 
Halliday (1998), and plusses galaxies from Kuntschner (1998).    
Also added are models with solar abundance ratios
of Vazdekis et al. (1996). The lines represent models of constant
metallicity (resp. Z=0.008, 0.02 and 0.05 from left to right), along which
age increases to the right. The line with thick symbols is the radial
profile of the Sombrero galaxy (see Section 5.1).}
\end{center}
\end{figure}

For the Sombrero galaxy the potential well is determined by both 
random and ordered, rotational motion ((v/$\sigma$)$_{\rm cen}$ $\sim$ 1,
Emsellem et al. 1996). Both $v$ and $\sigma_{cen}$ are larger than 
225 km/s, implying, as is indeed observed, that Mg/Fe has to be larger than
solar. Although the central region of the Sombrero is clearly 
dominated by a fast rotating disk, its Mg/Fe is much more similar
to the centres of other giant early-type galaxies (for which Mg/Fe $>$ 0)
than to large disks, like the disk of our Galaxy. If one believes 
that Mg/Fe only depends on the timescales of enrichment by SN type Ia and II, 
the fact that the 
range in observed central Mg/Fe amongst giant early-type galaxies is small
would then imply that the timescale on which stars formed there would be
independent on whether they are in a disk or not. Although we cannot measure
these timescales at present, this could sound rather implausible, and
it might be much more natural to assume that Mg/Fe depends on the IMF,
which in turn would be dependent on the environment through the stellar 
potential. At present we don't have any direct evidence that
the IMF varies in this way, but it is much easier to explain the 
observations in this way, as opposed to invoking differences in timescales
for enrichment by SN Ia and II.

The inner disk in the Sombrero is the only two-dimensional feature that
we have found in these three galaxies. This shows that the stellar populations
in the central regions of NGC~3379 and NGC~4472 are to a good approximation
homogeneous.

\subsection{Population Synthesis}

In this section we compare the measured line-strengths with 
the stellar population synthesis model of Paper I. This
comparison is very similar to the one performed in 
Paper II, apart from the fact that here we have a few extra
lines (H$\gamma$ and the Ca T), and that our spatial resolution 
here is higher. In Paper II we showed 
that very good fits for the three galaxies could be obtained with single-age, 
single-metallicity models, SSP's. Full evolutionary models provide solutions
that are very similar, with small ranges in age and metallicity. 
The best fits yield good agreement for all broad band optical and near 
infrared colors as well as for many important line indices.
Problems however are encountered when fitting Fe-dominated lines and 
the Ca4227 features.

To be able to use the higher spatial resolution of these observations, 
and to avoid too much duplication with Paper II, 
we have analyzed the light only in
an inner aperture with a diameter of 1.7$\arcsec$ centered on the
galaxy center. Best-fitting model solutions were found by minimizing 
a merit function {\it M} of the type described in Paper II. 
To calculate {\it M}, for each color or line index the square of the difference 
between the observed and synthetic color or index is divided by the estimated 
observational error and then multiplied by a certain assigned weight. 
To assign the weight to each index we have chosen to group them by 
their main contributing  element, and given the same global weight to each 
group. These groups have been defined are as follows: 
{\bf Balmer indices} (H$\beta$, H$\gamma_{A}$, H$\gamma_{F}$), 
{\bf Magnesium-dominated features} (Mg$_{1}$, Mg$_{2}$, Mg$_{b}$), 
{\bf Iron-dominated features} 
(Fe4383, Fe4531, Fe5015, Fe5270, Fe5335, \break Fe5406), {\bf Calcium-dominated features} 
(Ca4227 and the three lines of the Ca T) and {\bf other features} 
(G, Ca4455, C$_{2}$4668). The reason why Ca4455 is not included in the Ca 
group is because it is mainly a blend (Worthey et al. 1994). For that reason
we have always given a low weight to this line. 
To survey the solution space we have defined 4 weighting functions: 
in the first one the same weight is assigned  to each 
of the groups as a whole. To assess the influence of each group individually
three other weighting functions were defined where the weight for one
group was set to 0, while the other weights remained the same.  
All these weight functions are summarized in Table 5.

\begin{table}
\begin{center}
\begin{tabular}{l|ccccc}
\hline\hline
\multicolumn{5}{c}{Relative Weights used for the Merit Function}\\
\hline
H$\beta$     &1.0&1.0&1.0&1.0\\
H$\gamma_{A}$&1.0&1.0&1.0&1.0\\
H$\gamma_{F}$&1.0&1.0&1.0&1.0\\
\\
Mg$_{1}$     &1.0&0.0&1.0&1.0\\
Mg$_{2}$     &1.0&0.0&1.0&1.0\\
Mg$_{b}$     &1.0&0.0&1.0&1.0\\
\\
Fe4383       &0.5&0.5&0.0&0.5\\
Fe4531       &0.5&0.5&0.0&0.5\\
Fe5015       &0.5&0.5&0.0&0.5\\
Fe5270       &0.5&0.5&0.0&0.5\\
Fe5335       &0.5&0.5&0.0&0.5\\
Fe5406       &0.5&0.5&0.0&0.5\\
\\
Ca4227       &0.5&0.5&0.5&0.0\\
Ca{\sc II}(1)&0.5&0.5&0.5&0.0\\
Ca{\sc II}(2)&1.0&1.0&1.0&0.0\\
Ca{\sc II}(3)&1.0&1.0&1.0&0.0\\
\\
G            &1.0&1.0&1.0&1.0\\
Ca4455     &0.25&0.25&0.25&0.25\\
C$_{2}$4668  &1.0&1.0&1.0&1.0\\
\hline
\end{tabular}
\end{center}
\caption{Four different ways of defining the weights of the Merit 
Function}  
\end{table}

To fit the data, we have selected the models for three metallicities: Z=0.008,
Z=0.02 (solar) and\break Z=0.05, and interpolated the output linearly to obtain
results at Z=0.014 and 0.035. The age-range of the selected models was 1 to 17~Gyr.
We preferred to use the bimodal IMF defined in Paper I, 
which is similar to the Salpeter IMF, but has fewer  
stars with mass below 0.6M$_{\odot}$, since it was shown in Paper II
that this IMF provides slightly better fits to precisely the three galaxies
to be analyzed here. Four IMF slopes were considered: $\mu$ =  1.0, 1.3
(Salpeter), 1.7 and 2.3.

We will now discuss the fits for the galaxies individually. The best fits
were obtained for NGC~4472, giving a  metallicity of Z=0.05,
$\mu$=2.3  (steeper than Salpeter), and a most likely age of 8~Gyr. The same 
solution is found for all weight-functions,
except for the  case in which the Mg group is neglected, in which case 
an optimal age of 5~Gyr is found, with the same $\mu$ and metallicity.  
One should realize that we are not claiming here that with the Salpeter
IMF no reasonable solutions can be found for this galaxy. Using however the
bimodal IMF, which appears to be more realistic, at least for our Galaxy 
(Kroupa et al. 1993) it appears that
we need more stars around the mass where the IMF changes slope.
The fits show clearly that we cannot obtain at the same time
good fits for the Mg group
on one hand and the Fe+Ca groups on the other hand. This is the well known
problem that giant ellipticals are overabundant in Mg (see Worthey 1998 and
references therein for more details). Since the current theoretical models 
do not allow us to change the [$\alpha$/Fe] ratio, we will only continue 
with the uniform weight function for which no group has been neglected.
This fit is  summarized in Table 6. One sees that most of the features of the 
Ca group are much  less strong than
predicted by the model, and that these differences are clearly larger than the 
expected observational errors. This is the case for the lines of the Ca T, 
as well as for the Ca 4227 line, for which we already found in Paper 
II that it was much lower than expected.
We would like to warn the reader that the Ca T
predictions are based on an stellar library
(D{\'{\i}}az et al. 1989) which does not properly cover all the theoretical
atmospheric parameters.  It is reassuring that the present observational
values for the two main features of the Ca T are in agreement
with the observational results of Terlevich et al. (1990): while we measure a
total EW of 6.5\AA~ they measured 6.7\AA.  

The results for NGC~3379 are very similar as those for NGC~4472. 
Independent of the weight function the best solution is found for 
(Z=0.05, $\mu$=2.3, age=8~Gyr), except in the case in which the Mg group is
ignored, which gives the best solution for Z=0.035,
$\mu$=2.3, and age=8~Gyr.  The best fit with uniform weight function 
is also tabulated in Table 6. For this galaxy, the Ca
group features are lower than predicted, similar to NGC~4472.
In general the fact that the residuals in general are larger shows that
the quality of the obtained global fit is
considerably worse than the fit for NGC~4472.

For the central region of the Sombrero galaxy we always obtain Z=0.05,
$\mu$=2.3 and age=13~Gyr for all our weight functions except for the one
where we neglect the Mg group. Here the best fit is obtained for the same 
IMF slope, Z=0.035 and a slightly larger age (15~Gyr). 
It seems that possibly better fits could be obtained with even higher 
metallicities. This can however not be investigated because of the lack of
reliable theoretical models with Z $>$ 0.05. 
From the tabulated values we see that most of the features of the Ca
and Fe groups show residuals much larger than the  observed values and that the
Mg/Fe overabundance is very pronounced. Apart from this, we also see that while
the two H$\gamma$ indices are fairly well matched, this is not the case for
H$\beta$ which is much lower than any model prediction. From the spectra
it can be seen that this is the result of emission lines 'filling in' the
absorption line, and thereby reducing its equivalent width.
Overall this galaxy yielded fits that were qualitatively the worst of the three
galaxies.

To summarize our main, new results from this stellar population analysis:
The strengths of the lines of the Ca T are lower than expected,
confirming the result of Paper II purely on the basis of
the Ca 4227 line. We find that [Ca/Fe] = 0 or slightly
negative. This is peculiar, since Ca is thought to come primarily from
explosive and normal oxygen burning, and is expected to follow Mg in these
bright galaxies. Although our models seem to indicate that Ca is underabundant
with respect to Fe (a result also confirmed by Garc\'\i a-Vargas et al. 1998),
this is not certain, as explained in Idiart et al. (1997). Because of the fact
that the models of Paper I, like most currently available
models, assume that [Ca/Fe] and [Ca/Mg] = 0 for the input stars that
are used to calculate the fitting functions, the calculated integrated Ca 
T index will be too large if the stars in the input library that are used
are overabundant in Ca, and vice-versa. Idiart et al. have determined 
Ca abundances, as well as Fe and Mg abundances, for a medium-sized input 
library of about 100 stars, and using individual Ca/Fe ratios for those stars
they calculate models with much smaller values for the 
equivalent widths of the Ca T than Paper I or Garc\'\i
a-Vargas et al. (1998). Using models with solar abundance ratios we find that 
[Ca/Fe] for our three galaxies is about solar (see Fig.~10). 
Therefore, in spite of the fact that the Ca is an
$\alpha$-element, the Ca-dominated features do not track Mg,
but the Fe-dominated group. 

\onecolumn
\begin{table}
\footnotesize
\begin{center}
\begin{tabular}{l|r|rrr|rrr|rrr}
\hline\hline
\multicolumn{2}{c|}{}
&\multicolumn{3}{|c|}{NGC~4472}&\multicolumn{3}{|c|}{NGC~3379}&
\multicolumn{3}{|c}{NGC~4594}\\
\hline
\multicolumn{2}{c|}{}
&\multicolumn{3}{|c|}{(Z=0.05,$\mu$=2.3,Age=8Gyr)}&
\multicolumn{3}{|c|}{(Z=0.05,$\mu$=2.3,Age=8Gyr)}&
\multicolumn{3}{|c}{(Z=0.05,$\mu$=2.3,Age=13Gyr)}\\
Index        &Error&Observ.&Fit&Resid.&Observ.&Fit&Resid.&Observ.&Fit&Resid.\\
\hline
H$\beta$     &0.262& 1.325& 1.273& 0.2  & 1.048& 1.273& -0.9 & 0.374& 1.014& -2.4 \\
H$\gamma_{A}$&0.808&-6.674&-7.811& 1.4  &-6.747&-7.811&  1.3 &-7.891&-8.549& 0.8  \\
H$\gamma_{F}$&0.342&-1.771&-2.359& 1.7  &-2.091&-2.359&  0.8 &-2.638&-2.803& 0.5  \\
\\
Mg$_{1}$     &0.009& 0.165& 0.158& 0.8  & 0.171& 0.158& 1.4  & 0.190& 0.181& 1.0  \\
Mg$_{2}$     &0.010& 0.339& 0.331& 0.8  & 0.333& 0.331& 0.2  & 0.371& 0.365& 0.6  \\
Mg$_{b}$     &0.195& 5.160& 4.719& 2.3  & 5.159& 4.719& 2.3  & 5.141& 5.015& 0.7  \\
\\
Fe4383       &0.834& 6.178& 7.205& -1.2 & 5.692& 7.205& -1.8 & 6.470& 7.704& -1.5 \\
Fe4531       &0.455& 4.067& 4.181& -0.3 & 3.764& 4.181& -0.9 & 3.548& 4.433& -2.0 \\
Fe5015       &0.566& 6.315& 6.430& -0.2 & 5.401& 6.430& -1.8 & 3.446& 6.557& -5.5 \\
Fe5270       &0.317& 3.564& 3.627& -0.2 & 3.141& 3.627& -1.5 & 3.231& 3.818& -1.9 \\
Fe5335       &0.324& 2.666& 3.538& -2.7 & 2.701& 3.538& -2.6 & 2.757& 3.730& -3.0 \\
Fe5406       &0.228& 2.056& 2.293& -1.0 & 1.991& 2.293& -1.3 & 2.101& 2.439& -1.5 \\
\\
Ca4227       &0.220& 1.333& 2.072& -3.4  & 1.059& 2.072& -4.6 & 1.357& 2.362& -4.6 \\
Ca{\sc II}(1)&0.590& 1.413& 1.871& -0.8  & 1.178& 1.871& -1.2 & 1.510& 1.862& -0.6 \\
Ca{\sc II}(2)&0.581& 3.629& 4.496& -1.5 & 3.042& 4.496& -2.5 & 3.212& 4.403& -2.1 \\
Ca{\sc II}(3)&0.627& 2.855& 3.609& -1.2 & 2.368& 3.609& -2.0 & 2.894& 3.440& -0.9 \\
\\
G            &0.395& 6.048& 6.005& 0.1  & 6.136& 6.005& 0.3  & 5.966& 6.112& -0.4 \\
Ca4455       &0.288& 2.339& 2.133& 0.7  & 1.927& 2.133& -0.7 & 1.890& 2.264& -1.3  \\
C$_{2}$4668  &0.831& 8.737& 7.811& 1.1  & 8.318& 7.811& 0.6  & 9.578& 8.227& 1.6  \\
\hline
\end{tabular}
\end{center}
\caption{Fits using uniform weighting (first column in Table 5). 
For each galaxy we have tabulated age, metallicity and IMF slope (for a bimodal IMF). 
For each index we have
indicated the typical observational error  for our measurements, the fitted
values, and the residuals divided by the typical observational error. Note that
for NGC~4594 Fe 5015 and H$\beta$ are severely affected by emission lines.}
\end{table}

\twocolumn

\begin{figure}
\mbox{\epsfxsize=8cm  \epsfbox{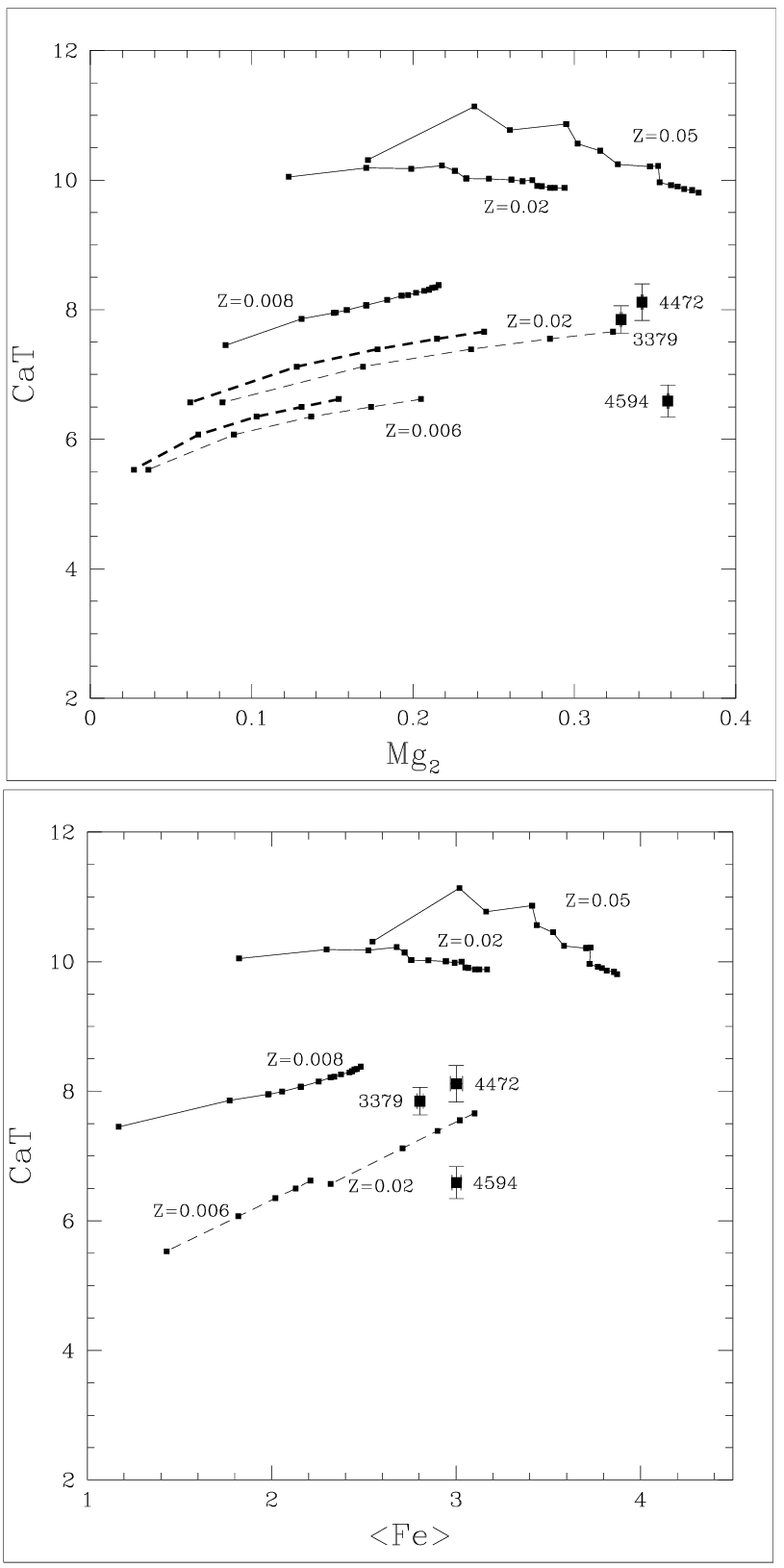}}
\caption{Models of the Ca T compared with the observations in the 
central aperture of 1.7$''$. The drawn lines are models of Paper II,
with metallicities of Z=0.008, 0.02 and 0.05, with age ranging from
1 to 17 Gyr from left to right. The dashed lines are the models of
Idiart et al. (1997) and Borges et al. (1995). 
In the upper figure the thick dashed lines indicate
models with solar [Mg/Fe], while for the thin dashed line [Mg/Fe]=0.30.}
\end{figure}

Combining the fits obtained here with those of Paper II obtained at
larger radii we can get an idea of the gradients in age, metallicity and  IMF
slope in these galaxies. To do this we summarize in Table 7 the fits at the 3
different positions. The table shows that the metallicity is the main  parameter
dominating the observed gradients. Although in the central regions 
of these galaxies a metallicity of more than 2 times solar is required, 
the metallicity decreases rapidly outward (being
below solar at 15$\arcsec$), except for NGC~4472, for which the gradient 
is shallower. In Paper II we
attributed this to the fact that this galaxy has a considerably larger
effective radius than the other two. The age on the other hand does not seem
to vary in an appreciable way with galactocentric radius. Only for NGC~3379 we
see a slight age gradient, being somewhat younger in its center. For our
understanding of these gradients it would be very useful if good observations
could be obtained further away from the center.

\begin{table}
\footnotesize
\begin{center}
\begin{tabular}{l|ccc}
\hline\hline
\multicolumn{4}{c}{Stellar Population Fits at various radii}\\
\hline
Galaxy& Center (Ap. 1.7$\arcsec$) & 5$\arcsec$ & 15$\arcsec$ \\
\hline
NGC~4472 & 0.05, 2.3, 8 & 0.04, 2.3, 10 & 0.03, 2.3 10 \\
NGC~3379 & 0.05, 2.3, 8 & 0.03, 1.5, 13 & 0.016, 2.3 12 \\
NGC~4594 & 0.05, 2.3, 13& 0.03, 2.3, 11 & 0.016, 2.0, 10\\
\hline
\end{tabular}
\end{center}
\caption{The best fits at three different radii, from this paper and 
Paper II. Tabulated are mean metallicity, IMF slope and age in Gyr. 
}  
\end{table}

\section{Conclusions}

We have observed the well-studied galaxies the Sombrero, NGC 3379 and NGC 4594.
We show that excellent two-dimensional absorption line strengths and stellar 
kinematics can be obtained with this instrument. We have extensively compared
our results with the current literature. A detailed comparison with 2D
multi-lenslet TIGER-spectroscopy of the Sombrero galaxy by Emsellem et al. (1996)
shows that we can
measure velocities and velocity dispersions to an accuracy of 10 km/s with
our current setup, or about 0.1 pixel. We reproduce well the absorption
line maps presented in that paper. An in-depth comparison with 
long-slit spectroscopy of Paper II shows in general good 
agreement between absorption line strength obtained using IFS and
long-slit spectroscopy. A problem in our absorption line maps is that
they contain peaks at random position which are not due to photon noise
but probably to continuum variations from fiber to fiber, possibly induced
by residual fringing or by variations in stress. 

We show that Mg/Fe is enhanced in the whole inner disk of the Sombrero galaxy.
[Mg/Fe] there is
larger than in the rest of the bulge. The large values of Mg/Fe in
the central disk are consistent with the centres of other early-type
galaxies, and not with large disks, like the disk of our Galaxy, where
[Mg/Fe] $\sim$ 0.  We confirm with this observation  
a recent result of Worthey (1998) that Mg/Fe is determined by  the central
kinetic energy, or escape velocity, of the stars, only, and not by the
formation time scale of the stars.
In galaxies with regions with
large escape velocities the IMF would have to
be skewed more towards high-mass stars, which then will favor SNe type II
as compared to SN type Ia, so that the Mg/Fe ratio will be enhanced.

We have obtained new observations of the Ca II IR triplet in the three galaxies,
and compared them with our model predictions. We find that the observations
of the Ca II IR triplet are lower than expected from the models. This result is
in agreement with Paper II, where we found that the Ca 4227 
line strength was lower than expected. The fact that it has been assumed in
the Vazdekis models (and most others) that [Ca/Fe] and [Ca/Mg] for the stars in
the stellar input libraries is solar, makes it difficult for 
our models to determine
the [Ca/Fe] ratio in our galaxies. Using however the models of Idiart et al.
(1997) who have taken these abundance ratios into account for individual stars,
we find that Ca approximately tracks Fe for our three galaxies.

We find that H$\gamma$ agrees well with predictions based on other lines,
including H$\beta$. Because of the fact that H$\beta$ 
is often severely contaminated by emission lines we confirm statements
by e.g. Kuntschner \& Davies (1998) that H$\gamma$ is often a very good alternative
to H$\beta$ when measuring e.g. ages of galaxies.

Combining the data presented in this paper with the results of
Paper II we find that the radial line strength gradients in 
the three galaxies are predominantly gradients in metallicity, in agreement
with earlier work (e.g. Davies et al. 1993).

\section*{Acknowledgements}  

We are grateful to Adolfo Garc\'\i a for his work developing 2D-FIS. This work has been
partially supported by the Direcci\'on General de Investigaci\'on Cient\'\i fica
y T\'ecnica (PB93-0658). We thank Martin Vogelaar for his programming help,
and the referee for useful comments that improved the paper.
The 4.2 m William Herschel Telescope is operated by the
Royal Greenwich Observatory, at the Spanish Observatorio del
Roque de los Muchachos of the Instituto de Astrof\'\i sica de Canarias. We thank all
the staff at the observatory for their kind support.

\newpage

\onecolumn

\end{document}